\documentclass[letterpaper, journal, 10pt]{IEEEtran}

\usepackage{etex}
\usepackage{cite}
\usepackage{nicefrac}
\usepackage{tikz}
\usepackage{pgfplots}
\usetikzlibrary{plotmarks,shapes,positioning,arrows,decorations.markings,fit,calc,patterns,mux}
\usepackage[usestackEOL]{stackengine} 

\usepackage[cmex10]{amsmath}
\usepackage{amssymb}
\usepackage[T1]{fontenc} 

\usepackage{bm}
\usepackage[varg]{txfonts}
\let\mathbb=\varmathbb
\DeclareSymbolFont{letters}{OML}{ztmcm}{m}{it}
\usepackage[font=footnotesize]{subfig}

\usepackage{tabularx,colortbl}
\usepackage{multirow}
\usepackage{dcolumn}
\usepackage{booktabs}
\newcolumntype{C}[1]{>{\centering\arraybackslash}p{#1}}

\newcommand{\sgn}[1]{\text{sgn}(#1)}

\input{figures/register.tikz} 
\input{figures/register_dotted.tikz} 

\usepackage{ifpdf}
\ifpdf
\usepackage[draft,pdfborder={0 0 0}]{hyperref}
\fi

\begin{document}

\title{Multi-mode Unrolled Architectures\\ for Polar Decoders}

\author{Pascal Giard, Gabi Sarkis, Claude Thibeault, and Warren J. Gross%
\thanks{P. Giard, G. Sarkis and W. J. Gross are with the Department of Electrical and Computer Engineering, McGill University, Montr\'eal, Qu\'ebec, Canada (e-mail: \{pascal.giard,gabi.sarkis\}@mail.mcgill.ca, warren.gross@mcgill.ca).}%
\thanks{C. Thibeault is with the Department of Electrical Engineering, \'Ecole de technologie sup\'erieure, Montr\'eal, Qu\'ebec, Canada (e-mail: claude.thibeault@etsmtl.ca).}}

\maketitle

\begin{abstract}
In this work, we present a family of architectures for polar decoders using a reduced-complexity successive-cancellation decoding algorithm that employs unrolling to achieve extremely high throughput values while retaining moderate implementation complexity. The resulting fully-unrolled, deeply-pipelined architecture is capable of achieving a coded throughput in excess of 1 Tbps on a 65 nm ASIC at 500 MHz---three orders of magnitude greater than current state-of-the-art polar decoders. However, unrolled decoders are built for a specific, fixed code. Therefore we also present a new method to enable the use of multiple code lengths and rates in a fully-unrolled polar decoder architecture. This method leads to a length- and rate-flexible decoder while retaining the very high speed typical to unrolled decoders. The resulting decoders can decode a master polar code of a given rate and length, and several shorter codes of different rates and lengths. We present results for two versions of a multi-mode decoder supporting eight and ten different polar codes, respectively. Both are capable of a peak throughput of 25.6 Gbps. For each decoder, the energy efficiency for the longest supported polar code is shown to be of 14.8~pJ/bit at 250~MHz and of 8.8~pJ/bit at 500~MHz.
\end{abstract}


\section{Introduction}
\IEEEPARstart{P}{olar} codes are gathering a lot of attention lately. They are error-correcting codes with an explicit construction that provably achieve the symmetric capacity of memoryless channels with a low-complexity decoding algorithm: successive cancellation (SC)~\cite{Arikan2009}. As SC proceeds bit-by-bit, hardware implementations suffered from low throughput and high latency \cite{Mishra2012,Leroux2012,Leroux2013,Raymond2014}. To overcome this, modified SC-based algorithms were proposed~\cite{Alamdar-Yazdi2011,Pamuk2013,Yuan2014,Sarkis_JSAC_2014,Li2014}. The first hardware implementation with a throughput greater than 1 Gbps was presented in \cite{Sarkis_JSAC_2014}. 

In \cite{Giard_IET_2015}, a fully-unrolled deeply-pipelined hardware architecture for polar decoders was proposed. Results showed a very high throughput, greater than 200~Gbps on FPGA. However, these architectures are built for a fixed polar code i.e. the code length or rate cannot be configured after designing the decoder. This is a major drawback for most modern wireless communication applications that largely benefit from the support of multiple code lengths and rates. Furthermore, a deeply-pipelined architecture causes the area to grow very fast with the frame size.

The goal of this paper is twofold. First, it is to generalize the unrolled architecture presented in \cite{Giard_IET_2015} into a family of architectures offering a flexible trade-off between throughput, area and energy efficiency. The (1024, 512) fully-unrolled deeply-pipelined polar decoder implementation of \cite{Giard_IET_2015} is significantly improved on all metrics. Second and most importantly, it is to show how an unrolled decoder built specifically for a polar code, of fixed length and rate, can be transformed into a multi-mode decoder supporting many codes of various lengths and rates. More specifically, we show how decoders for moderate-length polar codes contain decoders for many other shorter---but practical---polar codes of both high and low rates. The required hardware modifications are detailed, and ASIC synthesis and power estimations are provided for the 65 nm CMOS technology from TSMC. Results show a peak information throughput greater than 15 Gbps at 250 MHz in 4.29 mm$^2$ or greater than 20 Gbps at 500~MHz in 1.71~mm$^2$. Latency is of 2 $\mu$s and 650 ns for the former and latter.

The remainder of this paper starts with Section~\ref{sec:background} by briefly reviewing polar codes, their construction and their representation. Section~\ref{sec:ssc_and_fastssc} provides the necessary background on the Fast Simplified Successive-Cancellation (Fast-SSC) decoding algorithm. Section~\ref{sec:unrolled} describes the proposed family of unrolled hardware architectures. The concept, hardware modifications and other practical considerations related to the proposed multi-mode decoder are presented in Section~\ref{sec:concept}. Error-correction performance and implementation results for both dedicated and multi-mode decoders are provided in Section~\ref{sec:results}. Comparison against the fastest state-of-the-art polar decoder implementations in the literature is carried out in Section~\ref{sec:results} as well. Finally, a conclusion is drawn in Section~\ref{sec:conclusion}.

\section{Polar Codes}\label{sec:background}
\subsection{Construction}
Polar codes exploit the channel polarization phenomenon by which the probability of correctly estimating codeword bits tends to either 1 (completely reliable) or 0.5 (completely unreliable). These probabilities get closer to their limit as the code length increases when a recursive construction such as the one shown in Fig.~\ref{fig:pc16graph} is used, where $\oplus$ represents a modulo-2 addition (XOR). Under successive-cancellation decoding, polar codes were shown to achieve the symmetric capacity of memoryless channels as their code length $N \to \infty$ \cite{Arikan2009}.

An $(N, k)$ polar code has length $N$, carries $k$ information bits and is of rate $R=\nicefrac{k}{N}$. The other $N-k$ bits---frozen bits---are set to a predetermined value---usually zero---during the encoding process. 
The grayed $u_i$'s where $i \in \{0, 1, 2, 4\}$ on the left hand side of Fig.~\ref{fig:pc16graph} correspond to frozen bit locations of a (16, 12) polar code.

Depending on the type of channel and its conditions, the optimal location of the frozen bits varies and can be determined using the method described in \cite{Tal2011a} for example.

Encoding schemes for polar codes can be either non-systematic, as shown in Fig.~\ref{fig:pc16graph}, or systematic as discussed in \cite{Arikan2011}.
Systematic polar codes offer better bit-error rate (BER) than their non-systematic counterparts; while maintaining the same frame-error rate (FER). A low-complexity systematic encoding method was presented in \cite{Sarkis_JSAC_2014} and proven to be correct in \cite{Sarkis_TCOMM_2015}. In this work, we use systematic polar codes.

Both encoding types use the same generator matrix, and as this matrix is built recursively, so are polar codes i.e. a code of length $N$ is the concatenation of two codes of length $\nicefrac{N}{2}$. 

\subsection{Representation}
\begin{figure}
  \centering
  \hspace{-10pt}\resizebox{\columnwidth}{!}{\input{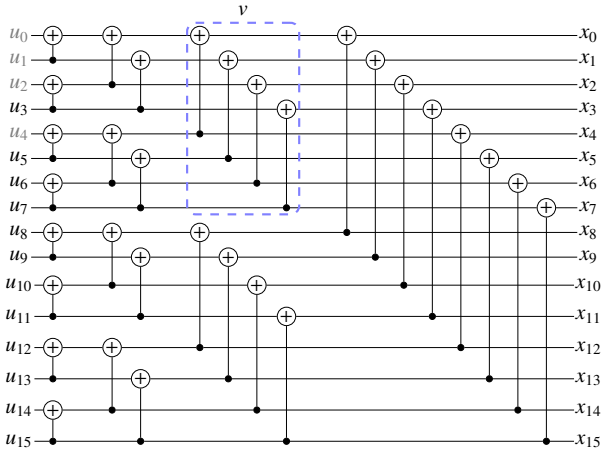}}
  \caption{Graph representation of a $(16, 12)$ polar code.}
  \label{fig:pc16graph}
\end{figure}

Fig.~\ref{fig:pc16graph} shows the graph representation of a $(16, 12)$ polar code where the blue-dashed-circled $v$ represents a concatenation of two codes of length 4, a $(4,1)$ polar code with a $(4,3)$ one, yielding an $(8,4)$ polar code.

\begin{figure}
  \centering
    \subfloat[SC]{\label{fig:tree-sc-16-12}
    \begin{tikzpicture}[every node/.style={font=\scriptsize}]
      \node[anchor=south west,inner sep=0] (image) at (0,0) {\includegraphics[width=0.7\columnwidth]{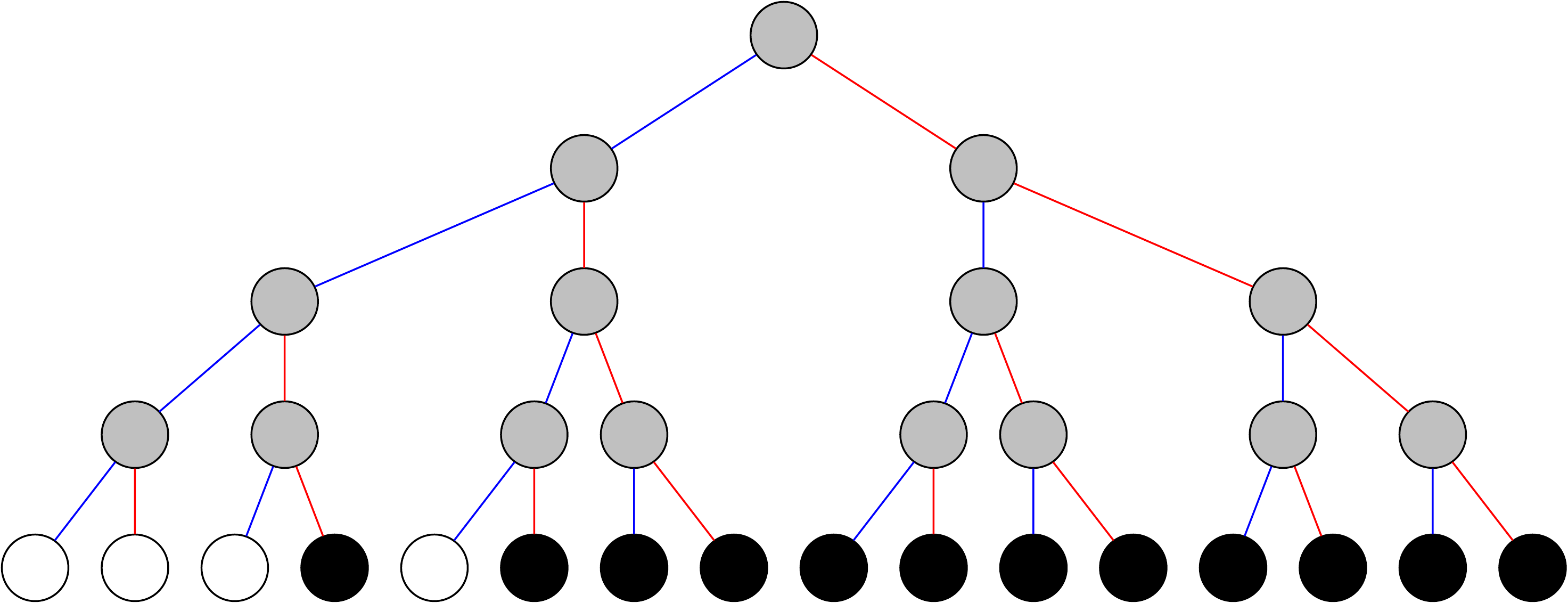}};
      \begin{scope}[x={(image.south east)},y={(image.north west)}]
        \node (v) at (.3725,.715) {$v$};
        \path[->,line width=0.65pt] ($(.47,.93)$) edge ($(.385,.78)$); 
        \path[->,line width=0.65pt] ($(.40,.73)$) edge ($(.485,.875)$); 
        \path[->,line width=0.65pt] ($(.345,.725)$) edge ($(.2,.56)$); 
        \path[->,line width=0.65pt] ($(.21,.5)$) edge ($(.355,.66)$); 
        \path[->,line width=0.65pt] ($(.36,.65)$) edge ($(.36,.56)$); 
        \path[->,line width=0.65pt] ($(.387,.56)$) edge ($(.387,.65)$); 
        \node (av) at (.405,.9) {$\alpha_v$};
        \node (bv) at (.465,.755) {$\beta_v$};
        \node (al) at (.25,.69) {$\alpha_l$};
        \node (bl) at (.275,.495) {$\beta_l$};
        \node (ar) at (.33,.55) {$\alpha_r$};
        \node (br) at (.42,.61) {$\beta_r$};
        \node (u0) at (.025,-.06) {$u_0$};
        \node (u1) at ($(u0)+(0.068,0)$) {$u_1$};
        \node (u2) at ($(u1)+(0.062,0)$) {$u_2$};
        \node (u3) at ($(u2)+(0.062,0)$) {$u_3$};
        \node (u4) at ($(u3)+(0.062,0)$) {$u_4$};
        \node (u5) at ($(u4)+(0.063,0)$) {$u_5$};
        \node (u6) at ($(u5)+(0.063,0)$) {$u_6$};
        \node (u7) at ($(u6)+(0.063,0)$) {$u_7$};
        \node (u8) at ($(u7)+(0.065,0)$) {$u_8$};
        \node (u9) at ($(u8)+(0.065,0)$) {$u_9$};
        \node (u10) at ($(u9)+(0.065,0)$) {$u_{10}$};
        \node (u11) at ($(u10)+(0.062,0)$) {$u_{11}$};
        \node (u12) at ($(u11)+(0.062,0)$) {$u_{12}$};
        \node (u13) at ($(u12)+(0.064,0)$) {$u_{13}$};
        \node (u14) at ($(u13)+(0.065,0)$) {$u_{14}$};
        \node (u15) at ($(u14)+(0.065,0)$) {$u_{15}$};
      \end{scope}
    \end{tikzpicture}}
  \subfloat[Fast-SSC]{\label{fig:tree-fastssc-16-12}\quad\resizebox{0.22\columnwidth}{!}{\rotatebox{90}{\definecolor{deepgreen}{RGB}{8, 130, 25}
\definecolor{mygray}{RGB}{192, 192, 192}

\begin{tikzpicture}[
        level/.style={level distance = 9mm},
        level 1/.style={sibling distance=11mm, edge from parent/.style={draw,blue,line width=0.5pt}},
        level 2/.style={sibling distance=15mm, edge from parent/.style={draw,blue,line width=0.5pt}},
        ]

\tikzset{
frozen/.style={draw=black,fill=white,minimum size=3mm,circle, inner sep=0},
fullspace/.style={draw=black,fill=black,minimum size=3mm,circle, inner sep = 0},
mixed/.style={draw=black,fill=mygray,minimum size=3mm,circle, inner sep = 0},
rep_mixed/.style={draw=black,pattern=north west lines,pattern color=deepgreen,minimum size=3mm,circle, inner sep = 0},
spc_mixed/.style={draw=black,pattern=crosshatch,pattern color=orange,minimum size=3mm,circle, inner sep = 0}
}

\tikzset{
parallel segment/.style={
   segment distance/.store in=\segDistance,
   segment pos/.store in=\segPos,
   segment length/.store in=\segLength,
   to path={
   ($(\tikztostart)!\segPos!(\tikztotarget)!\segLength/2!(\tikztostart)!\segDistance!90:(\tikztotarget)$) -- 
   ($(\tikztostart)!\segPos!(\tikztotarget)!\segLength/2!(\tikztotarget)!\segDistance!-90:(\tikztostart)$)  \tikztonodes
   }, 
   segment pos=.5,
   segment length=5ex,
   segment distance=-1mm,
},
}

\node[mixed] (3_0){} [grow=left]
	child {node[mixed] (2_0){\rotatebox{-90}{$v$}}
  	child {node[rep_mixed] (0_4){}
    }
		child {node[spc_mixed] (1_2){} edge from parent[red]
		}
	}
	child {node[fullspace] (2_1){} edge from parent[red]
	}
;

\draw[->,line width=0.65pt] (3_0) to[parallel segment,segment length=4ex] node[above right=-1.5mm] {\rotatebox{-90}{$\alpha_v$}} (2_0);
\draw[->,line width=0.65pt] (2_0) to[parallel segment] node[above right=-1.5mm] {\rotatebox{-90}{$\alpha_l$}} (0_4);
\draw[->,line width=0.65pt] (0_4) to[parallel segment] node[below left=-2.25mm] {\rotatebox{-90}{$\beta_l$}} (2_0);
\draw[->,line width=0.65pt] (2_0) to[parallel segment] node[above left=-2.25mm] {\rotatebox{-90}{$\alpha_r$}} (1_2);
\draw[->,line width=0.65pt] (1_2) to[parallel segment] node[below right=-2.25mm] {\rotatebox{-90}{$\beta_r$}} (2_0);
\draw[->,line width=0.65pt] (2_0) to[parallel segment,segment length=4ex] node[below left=-2.25mm] {\rotatebox{-90}{$\beta_v$}} (3_0);

\node at ($(0_4)-(.42,0)$) {\rotatebox{-90}{$u_0^3$}};
\node at ($(1_2)-(.42,0)$) {\rotatebox{-90}{$u_4^7$}};
\node at ($(2_1)-(.42,.1)$) {\rotatebox{-90}{$u_8^{15}$}};

\end{tikzpicture}}}}
  \vspace{-3pt}
  \caption{Decoder trees for SC (a) and Fast-SSC (b) decoding of a (16, 12) polar code.}
  \label{fig:tree-16-12}
\end{figure}
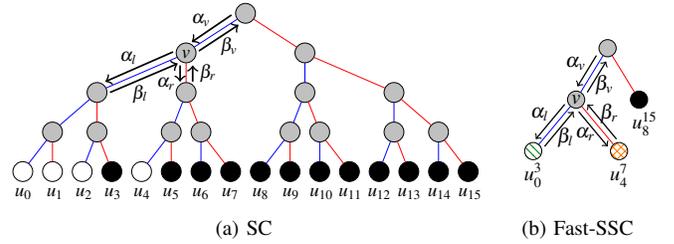

As polar codes are built recursively, it was proposed in \cite{Alamdar-Yazdi2011} to represent them as binary trees. Fig.~\ref{fig:tree-sc-16-12} illustrates such a representation, called decoder tree, equivalent to the graph of Fig.~\ref{fig:pc16graph}. In the decoder tree, white and black leaves represent frozen and information bits, respectively. Leaf nodes correspond to individual bits denoted $u_i$, where $0 \leq i <N$, and where the largest position index $i$ is on the right hand side of the tree. Moving up in the decoder tree corresponds to the concatenation of constituent codes. For example, the concatenation operation circled in blue in Fig.~\ref{fig:pc16graph} corresponds to the node labeled $v$ in Fig.~\ref{fig:tree-sc-16-12}.

The left-hand-side (LHS) and right-hand-side (RHS) subtrees rooted in the top node are polar codes of length $\nicefrac{N}{2}$. In the remainder of this paper, we designate the polar code, of length $N$, decoded by traversing the whole decoder tree as the \textit{master code} and the various codes of lengths smaller than $N$ as \textit{constituent codes}.

By definition, and like the master code, a constituent code of length $\nicefrac{N}{2}$ is in turn the concatenation of two polar codes of length $\nicefrac{N}{4}$, and so on until the leaf nodes are reached. As such, the decoding of a polar code of length $N$ can be seen as the decoding of two constituent codes of length $\nicefrac{N}{2}$, or of four constituent codes of length $\nicefrac{N}{4}$, etc. For example, and as shown in the graph representation of Fig.~\ref{fig:pc16graph}, but better seen in the decoder tree representation of Fig.~\ref{fig:tree-sc-16-12}, a master code of length $16$ is the concatenation of two constituent codes of length $8$, or of four constituent codes of length $4$, or of eight constituent codes of length $2$.

It should be noted that sibling constituent codes with the same parent node share a special relation. Let us consider the polar code (constituent code) of length $N_v=8$ taking root in $v$ as illustrated in Fig.~\ref{fig:tree-sc-16-12}, as the concatenation of two constituent codes of length $\nicefrac{N_v}{2}=4$. As that polar code gets decoded, the estimated bits $\beta_l$ from its LHS constituent code are required to compute the soft inputs $\alpha_r$ required to decode its RHS constituent code. Furthermore, once the estimated bits $\beta_r$ are obtained by decoding the RHS constituent code, they are combined with $\beta_l$ to form the bit-estimate vector $\beta_v$ for $v$.

\section{The Fast-SSC Decoding Algorithm}\label{sec:ssc_and_fastssc}
As mentioned above, a polar code is the concatenation of smaller constituent codes. Instead of using the SC algorithm on all constituent codes, the location of the frozen bits can be taken into account to use more efficient, lower complexity algorithms on some of these constituent codes~\cite{Alamdar-Yazdi2011, Sarkis_JSAC_2014}.

Fig.~\ref{fig:tree-fastssc-16-12} shows the decoder tree equivalent to Fig.~\ref{fig:tree-sc-16-12}, but when key parts of the Fast-SSC decoding algorithm~\cite{Sarkis_JSAC_2014} are used. The black node represents a rate-1 constituent code i.e. a polar code entirely composed of information bits. The green striped and orange cross-hatched nodes are repetition and single-parity-check (SPC) constituent codes, respectively. Gray nodes are codes of rate $0 < R < 1$.
It can be seen that Fast-SSC visits fewer nodes in the decoder tree, significantly decreasing the latency and increasing the throughput. It provides the same codeword estimates as SC though, hence offers the same error-correction performance.

While the proposed multi-mode unrolled decoders are independent of the decoding algorithm, we briefly go over the decoding operations mentioned in this paper.

\subsection*{Decoding Operations}\label{sec:fastssc}

Three functions are inherited from the original SC algorithm and log-likelihood ratios (LLRs) are used for the soft messages. Going down a left edge---colored blue in Fig.~\ref{fig:tree-16-12}---, $\alpha_l$ is calculated with the min-sum approximation \cite{Leroux2012}
\begin{equation}
\label{eqn:sc:f}
\begin{split}
\alpha_l[i] & = \sgn{\alpha_v[i] \cdot \alpha_v[i + \nicefrac{N_v}{2}]} \min(|\alpha_v[i]|, |\alpha_v[i + \nicefrac{N_v}{2}]|),
\end{split}
\end{equation}
for $0 \leq i < \nicefrac{N_v}{2}$, where $\alpha_v$ is the input to the node and $N_v$ the width of $\alpha_v$.
Going down a right edge---colored red in Fig.~\ref{fig:tree-16-12}---, $\alpha_r$ is calculated with
\begin{equation}
\label{eqn:sc:g}
\begin{split}
\alpha_r[i] & = \begin{cases}
\alpha_v[i + \nicefrac{N_v}{2}] + \alpha_v[i]\text{,} & \text{when } \beta_l[i] = 0;\\
\alpha_v[i + \nicefrac{N_v}{2}] - \alpha_v[i]\text{,} & \text{otherwise},
\end{cases}
\end{split}
\end{equation}
for $0 \leq i < \nicefrac{N_v}{2}$, where $\beta_l$ is the bit estimate from the LHS child.

Once a leaf node is reached, the bit estimate is set to zero when it corresponds to a frozen bit location. Otherwise, it is calculated by threshold detection on $\alpha_v$. Going back up a RHS edge the bit estimates from both children are combined to generate the node's bit-estimate vector
\begin{equation}
\label{eqn:sc:combine}
\beta_v[i] = \begin{cases}
\beta_l[i] \oplus \beta_r[i]\text{,} & \text{when } i < \nicefrac{N_v}{2};\\
\beta_r[i - \nicefrac{N_v}{2}]\text{,} & \text{when } \nicefrac{N_v}{2} \leq i < N_v \text{,}
\end{cases}
\end{equation}
where $\oplus$ is modulo-2 addition (XOR).

In \cite{Alamdar-Yazdi2011}, the Simplified SC (SSC) algorithm is introduced where decoder tree nodes are split into three categories: Rate-0, Rate-1, and Rate-$R$ nodes.

\subsubsection{Rate-0 Nodes} are subtrees whose leaf nodes all correspond to frozen bits. We do not need to use a decoding algorithm on such a subtree as the exact decision, by definition, is always the all-zero vector.

\subsubsection{Rate-1 Nodes} are subtrees where all leaf nodes carry information bits, none are frozen. The maximum-likelihood decoding rule for these nodes is to take a hard decision on the input LLRs:
\begin{equation}\label{eqn:ssc:info}
\beta_v[i] = \begin{cases}
  0, & \text{when } \alpha_v[i] \geq 0;\\
  1, & \text{otherwise,}
\end{cases}
\end{equation}
for $0 \leq i < N_v$. With a fixed-point representation, this operation amounts to copying the most significant bit of the input LLRs.

\subsubsection{Rate-$R$ Nodes} Lastly, Rate-$R$ nodes, where $0 < R < 1$, are subtrees such that leaf nodes are a mix of information and frozen bits. As shown in \cite{Sarkis_JSAC_2014}, instead of always using the SC or SSC algorithm, some Rate-$R$ nodes corresponding to specific frozen-bit locations can be decoded using algorithms with lower complexity and latency. The subset of nodes and operations from \cite{Sarkis_JSAC_2014} used in our proposed family of architectures are briefly reviewed in the following.

\subsubsection{$F$, $G$ and $G0R$ Operations}
The $F$ and $G$ operations are among the functions used in the conventional SC decoding algorithm and are calculated using \eqref{eqn:sc:f} and \eqref{eqn:sc:g}, respectively.

$G0R$ is a special case of the $G$ operation where the left child is a frozen node i.e. $\beta_l$ is known a priori to be the all-zero vector of length $\nicefrac{N_v}{2}$.

\subsubsection{$Combine$ and $C0R$ Operations}
As defined by \eqref{eqn:sc:combine}, the $Combine$ operation generates the bit estimate vector. A $C0R$ operation is a special case of the $Combine$ operation where the LHS constituent code, $\beta_l$, is a Rate-0 node.

\subsubsection{Repetition Node} In this node, all leaf nodes are frozen bits, with the exception of the node that corresponds to the most RHS leaf in a tree. At encoding time, the only information bit gets repeated over the $N_v$ outputs. The information bit can be estimated by using threshold detection over the sum of the input LLRs $\alpha_v$:
\[
\beta_v = \begin{cases}
0, & \text{when } \left(\sum_{i=0}^{N_v-1}{\alpha_v[i]}\right) \geq 0;\\
1, & \text{otherwise,}
\end{cases}
\]
where $\beta_v$ gets replicated $N_v$ times to create the bit-estimate vector.

\subsubsection{Single-parity-check (SPC) Node} An SPC node is a node such that all leaf nodes are information bits with the exception of the node at the least significant position (LHS leaf in a tree). To decode an SPC code, we start by calculating the parity of the input LLRs:
\[
\text{parity} = \bigoplus_{i = 0}^{N_v-1} \beta_v[i]\nonumber,
\text{ where }
\beta_v[i] = \begin{cases}
0, & \text{when } \alpha_v[i] \geq 0;\\
1, & \text{otherwise.}
\end{cases}
\]
The estimated bit vector is then generated by reusing the calculated $\beta_v$ above unless the parity constraint is not satisfied i.e. is different than zero. In that case, the estimated bit corresponding to the input with the smallest LLR magnitude is flipped:
\[
\beta_v[i] = \beta_v[i] \oplus 1, \text{where } i = \underset{j}{\arg\, \min}(|\alpha_v[j]|).\nonumber
\]

Our proposed decoders borrow from the Fast-SSC algorithm in that it uses specialized nodes and operations described above to reduce the decoding latency. However, the family of architectures we propose greatly differs from the processor-like architecture of \cite{Sarkis_JSAC_2014}. Moreover, \cite{Sarkis_JSAC_2014} proposes hybrid node types combining the ones above in order to further reduce the decoding latency. With the exception of the RepSPC node---a specialized node decoding a Repetition code concatenated with an SPC code---that is used in one of the implementations, we do not use those hybrid nodes in this paper.

\section{Unrolled Architectures}\label{sec:unrolled}

In an unrolled decoder, each and every operation required is instantiated so that data can flow through the decoder with minimal control.

The idea of fully unrolling a decoder has previously been applied to decoders for other families of error-correcting codes. Notably, in \cite{Schlafer2013,Wehn2015}, the authors propose a fully-unrolled deeply-pipelined decoder for an LDPC code. Polar codes are more suitable to unrolling as they do not feature a complex interleaver like LDPC codes.

\subsection{Deeply Pipelined}\label{sec:proposed_arch:deep}
In a deeply-pipelined architecture, a new frame is loaded into the decoder at every clock cycle. Therefore, a new estimated codeword is output at each clock cycle as each register is active at each rising edge of the clock (no enable signal required). In that architecture, at any point in time, there are as many frames being decoded as there are pipeline stages. This leads to a very high throughput at the cost of high memory requirements. Some pipeline stage paths do not contain any processing logic, only memory. They are added to ensure that the different messages remain synchronized. These added memories yield register chains, or SRAM blocks.

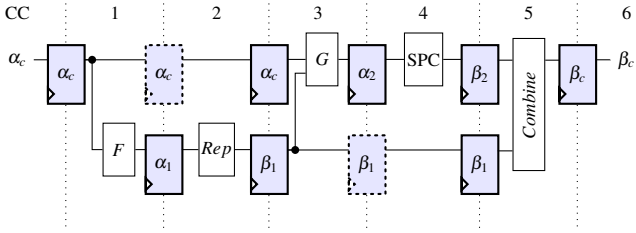
\begin{figure}[t]
  \centering
  \begin{tikzpicture}[font=\scriptsize,inner sep=1pt, minimum width=1.2em]

  \definecolor{deepgreen}{RGB}{8, 130, 25}

\tikzset{
branch/.style={fill,shape=circle,minimum size=3pt,inner sep=0pt},
block/.style={draw, rectangle, minimum height=2em},
spc/.style={draw, rectangle, minimum height=2em},
rep/.style={draw, rectangle, minimum height=2em},
comb/.style={draw, rectangle, minimum height=5em}
}

\node (ac) at (0.2,0) {$\alpha_c$};
\node at (0.2, 0.63) (cc) {CC};

\node[shape=reg] at ($(ac)+(0.65,-0.195)$) (REG1) {$\alpha_c$};

\node[shape=regdot] at ($(REG1)+(1.3,0)$) (REG2) {$\alpha_c$};
\node[block] at ($(REG1)+(0.7,-1.0)$) (F1) {$F$};
\node[shape=reg] at ($(F1)+(0.6,-0.2)$) (REG4) {$\alpha_1$};

\node[shape=reg] at ($(REG2)+(1.4,0)$) (REG3) {$\alpha_c$};
\node[rep] at ($(REG4)+(0.7,0.2)$) (Rep1) {$Rep$};
\node[shape=reg] at ($(Rep1)+(0.7,-0.2)$) (REG5) {$\beta_1$};

\node[block] at ($(REG3)+(0.7,0.2)$) (G1) {$G$};
\node[shape=reg] at ($(REG3)+(1.3,0)$) (REG6) {$\alpha_2$};
\node[shape=regdot] at ($(REG5)+(1.3,0)$) (REG7) {$\beta_1$};

\node[spc] at ($(REG6)+(0.75,0.2)$) (SPC1) {SPC};
\node[shape=reg] at ($(SPC1)+(0.75,-0.2)$) (REG8) {$\beta_2$};
\node[shape=reg] at ($(REG7)+(1.5,0)$) (REG9) {$\beta_1$};

\node[comb] at ($(REG8)+(0.65,-0.4)$) (Comb1) {\rotatebox{90}{$Combine$}};
\node[shape=reg] at ($(Comb1)+(0.65,0.4)$) (REG10) {$\beta_c$};

\node (bc) at ($(REG10)+(0.65,0.195)$) {$\beta_c$};
\draw let \p1 = (bc), \p2 = (cc) in node[coordinate] at (\x1, 0.63) (cc6) {6};
\node at (cc6) {6};

\draw[-] (ac.east) -- (REG1.D);

\draw[dotted] ($(REG1.Q)+(-0.25,0.75)$) -- ($(REG1.Q)+(-0.25,0.2)$) ($(REG1.Q)+(-0.25,-0.6)$) -- ($(REG1.Q)+(-0.25,-2.25)$);

\draw[-] (REG1.Q) -- (REG2.D);
\draw[-] (REG1.Q) -- ++(0.1,0) node[branch] {} -- ++(0,-0.45) |- (F1);
\draw[-] (F1) -| (REG4.D);

\draw[dotted] ($(REG2.Q)+(-0.25,0.75)$) -- ($(REG2.Q)+(-0.25,0.2)$) ($(REG2.Q)+(-0.25,-0.6)$) -- ($(REG4.Q)+(-0.25,0.2)$) ($(REG4.Q)+(-0.25,-0.6)$) -- ($(REG2.Q)+(-0.25,-2.25)$);

\draw[-] (REG4.Q) |- (Rep1) -| (REG5.D);

\draw[dotted] ($(REG3.Q)+(-0.25,0.75)$) -- ($(REG3.Q)+(-0.25,0.2)$) ($(REG3.Q)+(-0.25,-0.6)$) -- ($(REG5.Q)+(-0.25,0.2)$) ($(REG5.Q)+(-0.25,-0.6)$) -- ($(REG3.Q)+(-0.25,-2.25)$);

\draw[-] (REG2.Q) -- (REG3.D) (REG3.Q) |- (G1) -| (REG6.D);
\draw[-] (REG5.Q) -- ++(0.1,0) node[branch] {} |- ([yshift=-3mm]G1);
\draw[-] (REG5.Q) -- (REG7.D);

\draw[dotted] ($(REG6.Q)+(-0.25,0.75)$) -- ($(REG6.Q)+(-0.25,0.2)$) ($(REG6.Q)+(-0.25,-0.6)$) -- ($(REG7.Q)+(-0.25,0.2)$) ($(REG7.Q)+(-0.25,-0.6)$) -- ($(REG6.Q)+(-0.25,-2.25)$);

\draw[-] (REG6.Q) |- (SPC1) -| (REG8.D);
\draw[-] (REG7.Q) -- (REG9.D);

\draw[dotted] ($(REG8.Q)+(-0.25,0.75)$) -- ($(REG8.Q)+(-0.25,0.2)$) ($(REG8.Q)+(-0.25,-0.6)$) -- ($(REG9.Q)+(-0.25,0.2)$) ($(REG9.Q)+(-0.25,-0.6)$) -- ($(REG8.Q)+(-0.25,-2.25)$);

\draw[-] (REG8.Q) |- ([yshift=11mm]Comb1);
\draw[-] (REG9.Q) |- ([yshift=-11.5mm]Comb1);

\draw[-] (REG10.D) |- ([yshift=11mm]Comb1);

\draw[dotted] ($(REG10.Q)+(-0.25,0.75)$) -- ($(REG10.Q)+(-0.25,0.2)$) ($(REG10.Q)+(-0.25,-0.6)$) -- ($(REG10.Q)+(-0.25,-2.25)$);

\draw[-] (bc.west) |- (REG10.Q);

\draw let \p1 = (REG1), \p2 = (cc) in node[coordinate] at (\x1, \y2) (cc0) {};
\draw let \p1 = (REG4), \p2 = (cc) in node[coordinate] at (\x1, \y2) (cc1) {};
\draw let \p1 = (REG3), \p2 = (cc) in node[coordinate] at (\x1, \y2) (cc2) {};
\draw let \p1 = (REG6), \p2 = (cc) in node[coordinate] at (\x1, \y2) (cc3) {};
\draw let \p1 = (REG8), \p2 = (cc) in node[coordinate] at (\x1, \y2) (cc4) {};
\draw let \p1 = (REG10), \p2 = (cc) in node[coordinate] at (\x1, \y2) (cc5) {};

\node at ($(cc0)!0.5!(cc1)$) {1};
\node at ($(cc1)!0.5!(cc2)$) {2};
\node at ($(cc2)!0.5!(cc3)$) {3};
\node at ($(cc3)!0.5!(cc4)$) {4};
\node at ($(cc4)!0.5!(cc5)$) {5};

\end{tikzpicture}
  \caption{Fully-unrolled deeply-pipelined decoder for a (8, 4) polar code. Clock signals omitted for clarity.}
  \label{fig:unrolled_deeply_arch_8_4}
\end{figure}

Fig.~\ref{fig:unrolled_deeply_arch_8_4} shows a fully-unrolled and deeply-pipelined decoder for a $(8,4)$ polar code. The $\alpha$ and $\beta$ blocks illustrated in light blue are registers storing LLRs or bit estimates, respectively. White blocks are the functions described in Section~\ref{sec:fastssc} and dotted registers are regular registers but will be referred to in the next section. Among the registers, two are needed to retain the channel LLRs, denoted $\alpha_c$ in the figure, during the $2^{nd}$ and $3^{rd}$ clock cycles. Similarly, two registers have to be added for the persistence of the hard-decision vector $\beta_1$ over the $4^{th}$ and $5^{th}$ clock cycles. Such unrolled architectures for polar decoders were described in \cite{Giard_IET_2015}.

The information throughput can be defined as $\mathcal{P}fR$ bps, where $\mathcal{P}$ is the width of the output bus in bits, $f$ is the execution frequency in Hz and $R$ is the code rate. In this paper, $\mathcal{P}$ is assumed to be equal to the code length $N$. The decoding latency depends on the frozen bit locations and the constrained maximum width for all processing nodes, but is less than $N\log_2N$. In our experiments, with the operations and optimizations described below, the decoding latency never exceeded $\nicefrac{N}{2}$ clock cycles.

\subsection{Partially Pipelined}\label{sec:proposed_arch:partial}
In a deeply-pipelined architecture, a significant amount of memory is required for data persistence. That memory quickly increases with the code length $N$. Instead of loading a new frame into the decoder and estimating a new codeword at every cycle, we propose a compromise where the unrolled decoder can be partially pipelined to reduce the required memory. Let $\mathcal{I}$ be the initiation interval, where a new estimated codeword is output every $\mathcal{I}$ clock cycles. The case where $\mathcal{I}=1$ translates to a deeply-pipelined architecture. We note that the interval only affects the memory, not the computational elements, in the decoder.

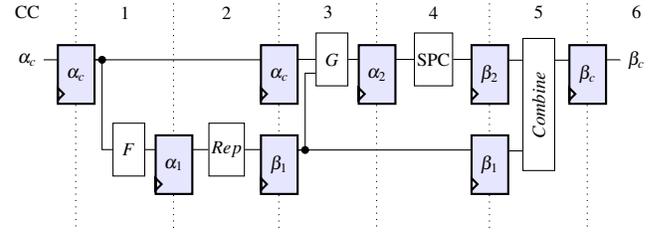
\begin{figure}[t]
  \centering
  \begin{tikzpicture}[font=\scriptsize,inner sep=1pt, minimum width=1.2em]

  \definecolor{deepgreen}{RGB}{8, 130, 25}

\tikzset{
branch/.style={fill,shape=circle,minimum size=3pt,inner sep=0pt},
block/.style={draw, rectangle, minimum height=2em},
spc/.style={draw, rectangle, minimum height=2em},
rep/.style={draw, rectangle, minimum height=2em},
comb/.style={draw, rectangle, minimum height=5em}
}

\node (ac) at (0.2,0) {$\alpha_c$};
\node at (0.2, 0.63) (cc) {CC};

\node[shape=reg] at ($(ac)+(0.65,-0.195)$) (REG1) {$\alpha_c$};

\node[block] at ($(REG1)+(0.7,-1.0)$) (F1) {$F$};
\node[shape=reg] at ($(F1)+(0.6,-0.2)$) (REG4) {$\alpha_1$};

\node[shape=reg] at ($(REG2)+(1.4,0)$) (REG3) {$\alpha_c$};
\node[rep] at ($(REG4)+(0.7,0.2)$) (Rep1) {$Rep$};
\node[shape=reg] at ($(Rep1)+(0.7,-0.2)$) (REG5) {$\beta_1$};

\node[block] at ($(REG3)+(0.7,0.2)$) (G1) {$G$};
\node[shape=reg] at ($(REG3)+(1.3,0)$) (REG6) {$\alpha_2$};

\node[spc] at ($(REG6)+(0.75,0.2)$) (SPC1) {SPC};
\node[shape=reg] at ($(SPC1)+(0.75,-0.2)$) (REG8) {$\beta_2$};
\node[shape=reg] at ($(REG7)+(1.5,0)$) (REG9) {$\beta_1$};

\node[comb] at ($(REG8)+(0.65,-0.4)$) (Comb1) {\rotatebox{90}{$Combine$}};
\node[shape=reg] at ($(Comb1)+(0.65,0.4)$) (REG10) {$\beta_c$};

\node (bc) at ($(REG10)+(0.65,0.195)$) {$\beta_c$};
\draw let \p1 = (bc), \p2 = (cc) in node[coordinate] at (\x1, 0.63) (cc6) {6};
\node at (cc6) {6};

\draw[-] (ac.east) -- (REG1.D);

\draw[dotted] ($(REG1.Q)+(-0.25,0.75)$) -- ($(REG1.Q)+(-0.25,0.2)$) ($(REG1.Q)+(-0.25,-0.6)$) -- ($(REG1.Q)+(-0.25,-2.25)$);

\draw[-] (REG1.Q) -- (REG3.D);
\draw[-] (REG1.Q) -- ++(0.1,0) node[branch] {} -- ++(0,-0.45) |- (F1);
\draw[-] (F1) -| (REG4.D);

\draw[dotted] ($(REG2.Q)+(-0.25,0.75)$) -- ($(REG4.Q)+(-0.25,0.2)$) ($(REG4.Q)+(-0.25,-0.6)$) -- ($(REG2.Q)+(-0.25,-2.25)$);

\draw[-] (REG4.Q) |- (Rep1) -| (REG5.D);

\draw[dotted] ($(REG3.Q)+(-0.25,0.75)$) -- ($(REG3.Q)+(-0.25,0.2)$) ($(REG3.Q)+(-0.25,-0.6)$) -- ($(REG5.Q)+(-0.25,0.2)$) ($(REG5.Q)+(-0.25,-0.6)$) -- ($(REG3.Q)+(-0.25,-2.25)$);

\draw[-] (REG3.D) (REG3.Q) |- (G1) -| (REG6.D);
\draw[-] (REG5.Q) -- ++(0.1,0) node[branch] {} |- ([yshift=-3mm]G1);
\draw[-] (REG5.Q) -- (REG9.D);

\draw[dotted] ($(REG6.Q)+(-0.25,0.75)$) -- ($(REG6.Q)+(-0.25,0.2)$) ($(REG6.Q)+(-0.25,-0.6)$) -- ($(REG6.Q)+(-0.25,-2.25)$);

\draw[-] (REG6.Q) |- (SPC1) -| (REG8.D);

\draw[dotted] ($(REG8.Q)+(-0.25,0.75)$) -- ($(REG8.Q)+(-0.25,0.2)$) ($(REG8.Q)+(-0.25,-0.6)$) -- ($(REG9.Q)+(-0.25,0.2)$) ($(REG9.Q)+(-0.25,-0.6)$) -- ($(REG8.Q)+(-0.25,-2.25)$);

\draw[-] (REG8.Q) |- ([yshift=11mm]Comb1);
\draw[-] (REG9.Q) |- ([yshift=-11.5mm]Comb1);

\draw[-] (REG10.D) |- ([yshift=11mm]Comb1);

\draw[dotted] ($(REG10.Q)+(-0.25,0.75)$) -- ($(REG10.Q)+(-0.25,0.2)$) ($(REG10.Q)+(-0.25,-0.6)$) -- ($(REG10.Q)+(-0.25,-2.25)$);

\draw[-] (bc.west) |- (REG10.Q);

\draw let \p1 = (REG1), \p2 = (cc) in node[coordinate] at (\x1, \y2) (cc0) {0};
\draw let \p1 = (REG4), \p2 = (cc) in node[coordinate] at (\x1, \y2) (cc1) {1};
\draw let \p1 = (REG3), \p2 = (cc) in node[coordinate] at (\x1, \y2) (cc2) {2};
\draw let \p1 = (REG6), \p2 = (cc) in node[coordinate] at (\x1, \y2) (cc3) {3};
\draw let \p1 = (REG8), \p2 = (cc) in node[coordinate] at (\x1, \y2) (cc4) {4};
\draw let \p1 = (REG10), \p2 = (cc) in node[coordinate] at (\x1, \y2) (cc5) {5};

\node at ($(cc0)!0.5!(cc1)$) {1};
\node at ($(cc1)!0.5!(cc2)$) {2};
\node at ($(cc2)!0.5!(cc3)$) {3};
\node at ($(cc3)!0.5!(cc4)$) {4};
\node at ($(cc4)!0.5!(cc5)$) {5};

\end{tikzpicture}
  \caption{Fully-unrolled partially-pipelined decoder for a (8, 4) polar code with $\mathcal{I}=2$. Clock signals omitted for clarity.}
  \label{fig:unrolled_partial_arch_8_4}
\end{figure}

Setting $\mathcal{I}$ $>$ 1 leads to a significant reduction in the memory requirements. An initiation interval of $\mathcal{I}$ translates to an effective required register chain length of $\lceil \nicefrac{L}{\mathcal{I}} \rceil$ instead of $L$, where $L$ is the length of the register chain. Using $\mathcal{I}=2$ leads to a $\sim50$\% reduction in the amount of memory required for that section of the circuit. This reduction applies to all register chains present in the decoder.
A partially-pipelined decoder with $\mathcal{I}=2$ can be obtained for a $(8,4)$ polar code by removing the dotted registers in Fig.~\ref{fig:unrolled_deeply_arch_8_4}, leading to the decoder of Fig.~\ref{fig:unrolled_partial_arch_8_4}.

The initiation interval $\mathcal{I}$ can be increased further in order to reduce the memory requirements, but only up to a certain limit. We call that limit the maximum initiation interval $\mathcal{I}_{\max}$, and its value depends on the decoder tree. By definition, the longest register chain in a fully-unrolled decoder is used to preserve the channel LLRs $\alpha_c$. Hence, the maximum initiation interval corresponds to the number of clock cycles required for the decoder to reach the last operation in the decoder tree that requires $\alpha_c$, $\text{G}_{N}$, the operation calculated when going down the right edge linking the root node to its right-hand-side child. Once that $\text{G}_{N}$ operation is completed, $\alpha_c$ is no longer needed and can be overwritten. 
As an example, consider the $(8,4)$ polar decoder illustrated in Fig.~\ref{fig:unrolled_partial_arch_8_4}. As soon as the switch to the right-hand side of the decoder tree occurs, i.e. when $G$ is traversed, the register containing the channel LLRs $\alpha_c$ can be updated with the LLRs for the new frame without affecting the remaining operations for the current frame. Thus the maximum initiation interval, $\mathcal{I}_{\max}$, for that decoder is 3.

The resulting coded and information throughput are
\begin{equation}\label{eqn:tp}
\mathcal{T}_C=\frac{N \cdot f}{\mathcal{I}}\text{~~~~and~~~~}\mathcal{T}_I=\frac{N \cdot f \cdot R}{\mathcal{I}},
\end{equation}
respectively, where $\mathcal{I}$ is the initiation interval. Note that this new definition can also be used for the deeply-pipelined architecture. The decoding latency remains unchanged compared to the deeply-pipelined architecture.

\begin{figure*}
  \centering
  \resizebox{2\columnwidth}{!}{\begin{tikzpicture}[font=\scriptsize,inner sep=1pt, minimum width=1.2em]

  \definecolor{deepgreen}{RGB}{8, 130, 25}

\tikzset{
branch/.style={fill,shape=circle,minimum size=3pt,inner sep=0pt},
block/.style={draw, rectangle, minimum height=2em},
spc/.style={draw, rectangle, minimum height=2em},
rep/.style={draw, rectangle, minimum height=2em},
comb/.style={draw, rectangle, minimum height=5em}
}

\node (ac) at (0.2,0) {$\alpha_0^{15}$};
\node at (0.2, 0.63) (cc) {CC};

\node[shape=reg] at ($(ac)+(0.775,-0.195)$) (REGAC1) {$\alpha_c$};

\phantom{\node[shape=reg] at ($(REGAC1)+(1.75,0)$) (REGAC2) {$\alpha_c$};}
\node[block] at ($(REGAC1)+(0.7,-0.65)$) (F1) {$F$};
\node[mux2,scale=0.55] at ($(F1)+(0.5,-0.275)$) (Mux1) {\large $m_1$};
\node (mode1_ain) at ($(Mux1.in1)-(0.4,0)$) {$\alpha_0^{7}$};
\node[shape=reg] at ($(Mux1)+(0.55,-0.19)$) (REG4) {$\alpha_1$};

\node[shape=reg] at ($(REGAC2)+(1.75,0)$) (REG3) {$\alpha_c$};
\phantom{\node[shape=reg] at ($(REG4)+(1.75,0)$) (REG6) {$\alpha_1$};}
\node[block] at ($(REG4)+(0.7,-0.85)$) (F2) {$F$};
\node[mux2,scale=0.55] at ($(F2)+(0.5,-0.275)$) (Mux2) {\large $m_2$};
\node (mode2_ain) at ($(Mux2.in1)-(0.4,0)$) {$\alpha_0^{3}$};
\node[shape=reg] at ($(Mux2)+(0.55,-0.19)$) (REG5) {$\alpha_2$};

\phantom{\node[shape=reg] at ($(REG3)+(1.3,0)$) (REG7) {$\alpha_c$};}
\node[shape=reg] at ($(REG6)+(1.3,0)$) (REG8) {$\alpha_1$};
\node[rep] at ($(REG5)+(0.6,0.2)$) (Rep1) {$Rep$};
\node[shape=reg] at ($(Rep1)+(0.70,-0.2)$) (REG9) {$\beta_1$};

\node[shape=reg] at ($(REG7)+(1.75,0)$) (REGAC5) {$\alpha_c$};
\node[block] at ($(REG8)+(0.7,0.2)$) (G1) {$G$};
\node[mux2,scale=0.55] at ($(G1)+(0.5,-0.275)$) (Mux3) {\large $m_3$};
\node (mode3_ain) at ($(Mux3.in1)-(0.4,0)$) {$\alpha_4^{7}$};
\node[shape=reg] at ($(Mux3)+(0.55,-0.19)$) (REGG1) {$\beta_2$};
\phantom{\node[shape=reg] at ($(REG9)+(1.75,0)$) (REG10) {$\beta_1$};}

\phantom{\node[shape=reg] at ($(REGAC5)+(1.35,0)$) (REGAC6) {$\alpha_c$};}
\node[spc] at ($(REGG1)+(0.6,0.2)$) (SPC1) {SPC};
\node[shape=reg] at ($(SPC1)+(0.75,-0.2)$) (REGSPC1) {$\beta_3$};
\node[shape=reg] at ($(REG10)+(1.35,0)$) (REG12) {$\beta_1$};

\node[shape=reg] at ($(REGAC6)+(1.4,0)$) (REGAC7) {$\alpha_c$};
\node[comb] at ($(REGSPC1)+(0.65,-0.4)$) (Comb1) {\rotatebox{90}{$Combine$}};
\node[shape=reg] at ($(Comb1)+(0.75,0.4)$) (REG14) {$\beta_4$};

\node[block] at ($(REGAC7)+(0.65,0.195)$) (G2) {$G$};
\node[block] at ($(G2)+(0.6,0.0)$) (Info3) {I};
\node[shape=reg] at ($(Info3)+(0.65,-0.2)$) (REG15) {$\beta_5$};
\phantom{\node[shape=reg] at ($(REG14)+(1.9,0)$) (REG16) {$\beta_4$};}

\node[comb,minimum height=5.6em] at ($(REG15)+(0.65,-0.5)$) (Comb2) {\rotatebox{90}{$Combine$}};
\node[mux2,scale=0.55] at ($(Comb2)+(0.4,-2.35)$) (Mux4) {\large $m_4$};
\node[block] at ($(Mux4.in1)+(-0.4,0)$) (Concat1) {$\&$};
\node[mux2,scale=0.55] at ($(Comb2)+(1.05,-0.9)$) (Mux5) {\large $m_5$};
\node[block] at ($(Comb2)+(1.7,0.5)$) (Concat2) {$\&$};
\node[shape=reg] at ($(Concat2)+(0.6,0.)$) (REG17) {$\beta_c$};

\node (bc) at ($(REG17)+(0.7,0.195)$) {$\beta_0^{15}$};
\draw let \p1 = (bc), \p2 = (cc) in node[coordinate] at (\x1, 0.63) (cc9) {9};
\node at (cc9) {9};

\draw[-] (ac.east) -- (REGAC1.D);

\draw[dotted] ($(REGAC1.Q)+(-0.25,0.5)$) -- ($(REGAC1.Q)+(-0.25,0.2)$) ($(REGAC1.Q)+(-0.25,-0.6)$) -- ($(REGAC1.Q)+(-0.25,-3.9)$);

\draw[-] (REGAC1.Q) -- (REGAC2.D) -- (REGAC2.Q);
\draw[-] (REGAC1.Q) -- ++(0.1,0) node[branch] {} -- ++(0,-0.45) |- (F1) -- (Mux1.in0);
\draw[-] (mode1_ain) -- (Mux1.in1);
\draw[-] (Mux1.out) |- (REG4.D);

\draw[dotted] ($(REGAC2.Q)+(-0.25,0.5)$) -- ($(REGAC2.Q)+(-0.25,0.2)$) -- ($(REGAC2.Q)+(-0.25,-0.6)$) -- ($(REG4.Q)+(-0.25,0.2)$) ($(REG4.Q)+(-0.25,-0.6)$) -- ($(REGAC2.Q)+(-0.25,-3.9)$);

\draw[-] (REGAC2.Q) -- (REG3.D);
\draw[-] (REG4.Q) -- (REG6.D) -- (REG6.Q);
\draw[-] (REG4.Q) -- ++(0.1,0) node[branch] {} -- ++(0,-0.45) |- (F2) -| (Mux2.in0);
\draw[-] (mode2_ain) -- (Mux2.in1);
\draw[-] (Mux2.out) |- (REG5.D);

\draw[dotted] ($(REG3.Q)+(-0.25,0.5)$) -- ($(REG3.Q)+(-0.25,0.2)$) ($(REG3.Q)+(-0.25,-0.6)$) -- ($(REG6.Q)+(-0.25,0.2)$) -- ($(REG6.Q)+(-0.25,-0.6)$) -- ($(REG5.Q)+(-0.25,0.2)$) ($(REG5.Q)+(-0.25,-0.6)$) -- ($(REG3.Q)+(-0.25,-3.9)$);

\draw[-] (REG3.Q) -- (REG7.D) -- (REG7.Q);
\draw[-] (REG6.Q) -- (REG8.D);
\draw[-] (REG5.Q) |- (Rep1) -| (REG9.D);

\draw[dotted] ($(REG7.Q)+(-0.25,0.5)$) -- ($(REG7.Q)+(-0.25,0.2)$) -- ($(REG7.Q)+(-0.25,-0.6)$) -- ($(REG8.Q)+(-0.25,0.2)$) ($(REG8.Q)+(-0.25,-0.6)$) -- ($(REG9.Q)+(-0.25,0.2)$) ($(REG9.Q)+(-0.25,-0.6)$) -- ($(REG7.Q)+(-0.25,-3.9)$);

\draw[-] (REG7.Q) -- (REGAC5.D);
\draw[-] (REG8.Q) |- (G1) -| (Mux3.in0) (Mux3.out) -| (REGG1.D) (REGG1.Q) |- (SPC1) -| (REGSPC1.D);
\draw[-] (mode3_ain) -- (Mux3.in1);

\draw[-] (REG9.Q) |- (REG10.D) -- (REG10.Q) |- (REG12.D);
\draw[-] (REG9.Q) ++(0.15,0) node[branch] {} |- ([yshift=-3.0mm]G1);

\draw[dotted] ($(REGAC5.Q)+(-0.25,0.5)$) -- ($(REGAC5.Q)+(-0.25,0.2)$) ($(REGAC5.Q)+(-0.25,-0.6)$) -- ($(REGG1.Q)+(-0.25,0.2)$) ($(REGG1.Q)+(-0.25,-0.6)$) -- ($(REG10.Q)+(-0.25,0.2)$) -- ($(REG10.Q)+(-0.25,-0.6)$) -- ($(REGAC5.Q)+(-0.25,-3.9)$);

\draw[-] (REGAC5.Q) -- (REGAC6.D) -- (REGAC6.Q);
\draw[-] (REGSPC1.Q) |- ([yshift=13mm]Comb1);
\draw[-] (REG12.Q) |- ([yshift=-11mm]Comb1);
\draw[-] ([yshift=9mm]Comb1) -| ([xshift=-1.8mm]REG14.D) -- (REG14.D);
\draw[-] ([xshift=-1.8mm]REG14.D) |- (Mux4.in0);
\fill[black] ([xshift=-1.8mm]REG14.D) circle (1.5pt);

\draw[dotted] ($(REGAC6.Q)+(-0.25,0.5)$) -- ($(REGAC6.Q)+(-0.25,0.2)$) -- ($(REGAC6.Q)+(-0.25,-0.6)$) -- ($(REGSPC1.Q)+(-0.25,0.2)$) ($(REGSPC1.Q)+(-0.25,-0.6)$) -- ($(REG12.Q)+(-0.25,0.2)$) ($(REG12.Q)+(-0.25,-0.6)$) -- ($(REGAC6.Q)+(-0.25,-3.9)$);

\draw[dotted] ($(REGAC7.Q)+(-0.25,0.5)$) -- ($(REGAC7.Q)+(-0.25,0.2)$) ($(REGAC7.Q)+(-0.25,-0.6)$) -- ($(REG14.Q)+(-0.25,0.2)$) ($(REG14.Q)+(-0.25,-0.6)$) -- ($(REGAC7.Q)+(-0.25,-3.9)$);

\draw[-] (REGAC6.Q) -- (REGAC7.D) (REGAC7.Q) --(G2) -- (Info3) -| (REG15.D);
\draw[-] (REG14.Q) ++(0.1,0) node[branch] {} |- ([yshift=-2.5mm]G2);
\draw[-] (REG14.Q) -- (REG16.D);

\draw[dotted] ($(REG15.Q)+(-0.25,0.5)$) -- ($(REG15.Q)+(-0.25,0.2)$) ($(REG15.Q)+(-0.25,-0.6)$) -- ($(REG16.Q)+(-0.25,0.2)$) -- ($(REG16.Q)+(-0.25,-0.6)$) -- ($(REG15.Q)+(-0.25,-3.9)$);

\draw[-] (REG15.Q) |- ([yshift=13mm]Comb2);
\draw[-] (REG16.D) -- (REG16.Q) |- ([yshift=-12.7mm]Comb2) ([yshift=-26mm]Comb2) -| (Mux5.in0);

\draw[-] (Concat1) |- (Mux4.in1);
\draw[-] (SPC1) ++(0.35,0) |- ([yshift=35mm]Concat1);
\draw[-] (Rep1) ++(0.35,0) |- ([yshift=-65mm]Concat1);
\fill[black] ($(SPC1)+(0.35,0)$) circle (1.5pt);
\fill[black] ($(Rep1)+(0.35,0)$) circle (1.5pt);

\draw[-] (Mux4.out) -- ++(0.15,0) |- (Mux5.in1);
\draw[-] (Comb2) ++(0.22,0.65) |- ([yshift=10mm]Concat2) (Concat2) ++(0.21,0.2) -| (REG17.D);
\draw[-] (Mux5.out) -- ++(.15,0) |- ([yshift=-1em]Concat2);
\node (comb2out1) at ($(Comb2)+(0.51,0.75)$) {\tiny $[15..8]$};
\node (comb2out1) at ($(Comb2)+(0.47,-0.76)$) {\tiny $[7..0]$};

\draw[dotted] ($(REG17.Q)+(-0.25,0.5)$) -- ($(REG17.Q)+(-0.25,0.2)$) ($(REG17.Q)+(-0.25,-0.6)$) -- ($(REG17.Q)+(-0.25,-3.9)$);

\draw[-] (REG17.Q) |- (bc.west);

\draw let \p1 = (REGAC1), \p2 = (cc) in node[coordinate] at (\x1, \y2) (cc0) {0};
\draw let \p1 = (REGAC2), \p2 = (cc) in node[coordinate] at (\x1, \y2) (cc1) {1};
\draw let \p1 = (REG3), \p2 = (cc) in node[coordinate] at (\x1, \y2)   (cc2) {2};
\draw let \p1 = (REG8), \p2 = (cc) in node[coordinate] at (\x1, \y2)   (cc3) {3};
\draw let \p1 = (REGAC5), \p2 = (cc) in node[coordinate] at (\x1, \y2) (cc4) {4};
\draw let \p1 = (REGSPC1), \p2 = (cc) in node[coordinate] at (\x1, \y2)(cc5) {5};
\draw let \p1 = (REGAC7), \p2 = (cc) in node[coordinate] at (\x1, \y2) (cc6) {6};
\draw let \p1 = (REG15), \p2 = (cc) in node[coordinate] at (\x1, \y2)  (cc7) {7};
\draw let \p1 = (REG17), \p2 = (cc) in node[coordinate] at (\x1, \y2)  (cc8) {8};

\node at ($(cc0)!0.5!(cc1)$) {1};
\node at ($(cc1)!0.5!(cc2)$) {2};
\node at ($(cc2)!0.5!(cc3)$) {3};
\node at ($(cc3)!0.5!(cc4)$) {4};
\node at ($(cc4)!0.5!(cc5)$) {5};
\node at ($(cc5)!0.5!(cc6)$) {6};
\node at ($(cc6)!0.5!(cc7)$) {7};
\node at ($(cc7)!0.5!(cc8)$) {8};

\end{tikzpicture}}
  \caption{Unrolled partially-pipelined decoder for a $(16, 12)$ polar code with initiation interval $\mathcal{I} = 2$. Clock, flip-flop enable and multiplexer select signals are omitted for clarity.}
  \vspace{-15pt}
  \label{fig:unrolled-impl}
\end{figure*}
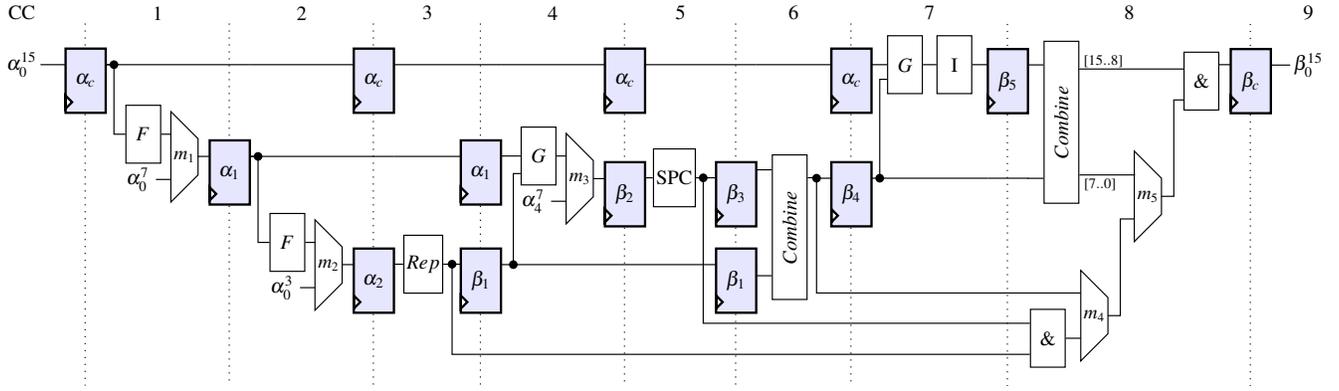

Fig.~\ref{fig:unrolled-impl} shows a fully-unrolled partially-pipelined decoder with an initiation interval $\mathcal{I}=2$ for the $(16, 12)$ polar code of Fig.~\ref{fig:tree-fastssc-16-12}. Some control and routing logic was added to make it multi-mode as detailed in the next section. The ``$\&$'' blocks are bit-vector joining operators. 

The partially-pipelined architecture requires a more elaborate controller than the deeply-pipelined architecture. For both fully- and partially-pipelined architectures, the controller generates a done signal to indicate that a new estimated codeword is available at the output. For the partially-pipelined architecture, the controller also contains a counter with maximum value of $(\mathcal{I}-1)$ which generates the $\mathcal{I}$ enable signals for the registers. An enable signal is asserted only when the counter reaches its value, in $[0, \mathcal{I}-1]$, otherwise it remains deasserted. Each register uses an enable signal corresponding to its location in the pipeline modulo $\mathcal{I}$. As an example, let us consider the decoder of Fig.~\ref{fig:unrolled-impl}, i.e. $\mathcal{I}$ is set to 2. In that example, two enable signals are created and a simple counter alternates between 0 and 1. The registers storing the channel LLRs $\alpha_c$ are enabled when the counter is equal to 0 because their input resides on the even (0, 2, 4 and 6) stages of the pipeline. On the other hand, the two registers holding the $\alpha_1$ LLRs are enabled when the counter is equal to 1 because their inputs are on odd (1 and 3) stages. The other registers follow the same rule.

The required memory resources could be further reduced by performing the decoding operations in a combinational manner, i.e. by removing all the registers except the ones labeled $\alpha_c$ and $\beta_c$, as in \cite{Dizdar2015}. However, the resulting reachable frequency is too low for the desired throughput level.

\subsection{Replacing Register Chains with SRAM Blocks}\label{sect:impl:sram}
As the code length $N$ grows, long register chains start to appear in the decoder, especially with a smaller $\mathcal{I}$. In order to reduce the number of registers required, register chains can be converted into SRAM blocks.

Consider the register chain of length 4 used for the persistence of the channel LLRs $\alpha_c$ in the fully-unrolled partially-pipelined $(16,12)$ decoder shown in top row of Fig.~\ref{fig:unrolled-impl}. Preserving the first register, the remaining 3 registers in that chain can be replaced by a dual-port SRAM block with a width of $16Q$ bits---$Q$ is the number quantization bits---and depth of 3 along with a controller to generate the appropriate read and write addresses. Similar to a circular buffer, if the addresses are generated to increase every clock cycle, the write address is set to be one position ahead of the read address.

SRAM blocks can replace register chains in a deeply-pipelined architecture as well. In both architectures, the SRAM block depth has to be equal or greater than the register chain length minus one.

\section{Multi-mode Unrolled Decoders}\label{sec:concept}

It can be noted that an unrolled decoder for a polar code of length $N$ is composed of unrolled decoders for two polar codes of length $\nicefrac{N}{2}$, which are each composed of unrolled decoders for two polar codes of length $\nicefrac{N}{4}$, and so on. Thus, by adding some control and routing logic, it is possible to directly feed and read data from the unrolled decoders for constituent codes of length smaller than $N$. The end result is a multi-mode decoder supporting frames of various lengths and code rates.

\subsection{Hardware Modifications to the Unrolled Decoders}
Consider the decoder tree shown in Fig.~\ref{fig:tree-fastssc-16-12} along with its unrolled implementation as illustrated in Fig.~\ref{fig:unrolled-impl}.
In Fig.~\ref{fig:tree-fastssc-16-12}, the constituent code taking root in $v$ is an $(8,4)$ polar code.
Its corresponding decoder can be directly employed by placing the 8 channels LLRs into $\alpha_0^7$ and by selecting the bottom input of the multiplexer $m_1$ illustrated in Fig.~\ref{fig:unrolled-impl}.
Its estimated codeword is retrieved from reading the output of the $Combine$ block feeding the $\beta_4$ register i.e. by selecting the top and bottom inputs from $m_4$ and $m_5$, respectively, and by reading the 8 least-significant bits from $\beta_0^{15}$.
Similarly, still in Fig.~\ref{fig:unrolled-impl}, the decoders for the repetition and SPC constituent codes can be fed via the $m_2$ and $m_3$ multiplexers and their output eventually recovered from the output of the $Rep$ and SPC blocks, respectively.

Although not illustrated in Figs.~\ref{fig:unrolled_deeply_arch_8_4}, \ref{fig:unrolled_partial_arch_8_4} or \ref{fig:unrolled-impl}, the proposed unrolled decoders feature a minimal controller. While not mandatory, the functionality of this controller is altered to better accommodate the use of multiple polar codes. Two look-up tables (LUTs) are added. One LUT stores the decoding latency, in clock cycles, of each code. It serves as a stopping criteria to generate the done signal. The other LUT stores the clock cycle ``value'' $i_{\text{start}}$ at which the enable-signal generator circuit should start.
Each non-master code may start at a value $(i_\text{start} \mod \mathcal{I}) \neq 0$. In such cases, using the unaltered controller would result in the waste of $(i_\text{start} \mod \mathcal{I})$ clock cycles. It can be significant for short codes, especially with large values of $\mathcal{I}$.
For example, without these changes, for the implementation with a master code of length 1024 and $\mathcal{I}=20$ presented in Section~\ref{sec:results} below, the latency for the $(128, 96)$ polar code would increase by 20\% as $(i_\text{start} \mod \mathcal{I}) = 17$ and the decoding latency is of 82 clock cycles.
 
Lastly, the modified controller also generates the multiplexer select signals, allowing proper data routing, based on the selected mode.

\subsection{On the Construction of the Master Code}\label{sec:assembled}
Conventional approaches construct polar codes for a given channel type and condition. In this work, many of the constituent codes contained within a master code are not only used internally to detect and correct errors, they are used separately as well. Therefore, we propose to assemble a master code using two optimized constituent codes in order to increase the number of optimized polar codes available. Doing so, the number of information bits, or the code rate, of the second largest supported codes can be selected. In the following, a master code of length 2048 is constructed by concatenating two constituent codes of length 1024. The LHS and RHS constituent codes are chosen to have a rate of $\nicefrac{1}{2}$ and of $\nicefrac{5}{6}$, respectively. As a result, the assembled master code has rate $\nicefrac{2}{3}$. The location of the frozen bits in the master code is dictated by its constituent codes. Note that the constituent code with the lowest rate is put on the left---and the one with the highest rate on the right---to minimize the coding loss associated with a non-optimized polar code.

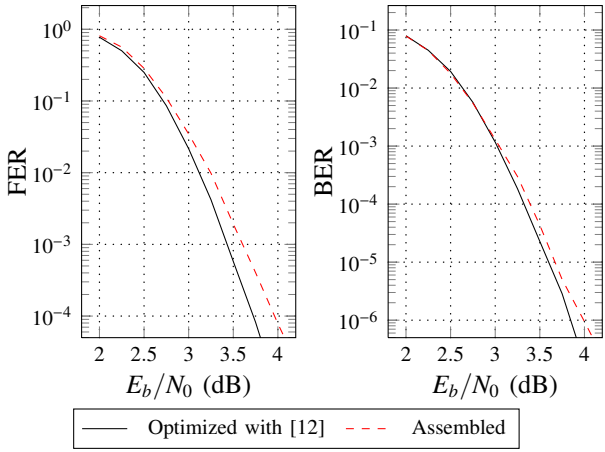
\begin{figure}
  \centering
  \begin{tikzpicture}

  \pgfplotsset{
    grid style = {
      dash pattern = on 0.05mm off 1mm,
      line cap = round,
      black,
      line width = 0.5pt
    },
    label style = {font=\fontsize{10pt}{7.2}\selectfont},
    tick label style = {font=\fontsize{8pt}{7.2}\selectfont}
  }

  \begin{semilogyaxis}[%
    xlabel=$E_b/N_0$ (dB),%
    xlabel style={yshift=0.5em},%
    xtick={2,2.5,3,3.5,4},%
    ymin=5e-5,
    ylabel=FER, ylabel style={yshift=-1.05em},%
    width=0.5\columnwidth, height=6.0cm, grid=major,%
    legend style={
      anchor={center},
      cells={anchor=west},
      column sep= 2mm,
      font=\fontsize{8pt}{7.2}\selectfont,
    },
    legend to name=perf-3-legend,
    legend columns=4,
    mark size=3.0pt]

    \addplot[color=black,solid] table[x=ebn0_db,y=FER] {data/2048.1365.s0.546.csv};
    \addlegendentry{Optimized with \cite{Tal2011a}}

    \addplot[color=red,dashed] table[x=ebn0_db,y=FER] {data/merged_n1k_awgn_s0.75+n1k_awgn_s0.436.float.csv};
    \addlegendentry{Assembled}

  \end{semilogyaxis}
\end{tikzpicture}
\begin{tikzpicture}

  \pgfplotsset{
    grid style = {
      dash pattern = on 0.05mm off 1mm,
      line cap = round,
      black,
      line width = 0.5pt
    },
    label style = {font=\fontsize{10pt}{7.2}\selectfont},
    tick label style = {font=\fontsize{8pt}{7.2}\selectfont}
  }

  \begin{semilogyaxis}[%
    xlabel=$E_b/N_0$ (dB),%
    xlabel style={yshift=0.5em},%
    xtick={2,2.5,3,3.5,4},%
    ymin=5e-7,
    ylabel=BER, ylabel style={yshift=-1.05em},%
    width=0.5\columnwidth, height=6.0cm, grid=major,%
    mark size=3.0pt]

    \addplot[color=black,solid] table[x=ebn0_db,y=BER] {data/2048.1365.s0.546.csv};

    \addplot[color=red,dashed] table[x=ebn0_db,y=BER] {data/merged_n1k_awgn_s0.75+n1k_awgn_s0.436.float.csv};

  \end{semilogyaxis}
\end{tikzpicture}
\\
\ref{perf-3-legend}
  \caption{Error-correction performance of two $(2048, 1365)$ polar codes with different constructions.}
  \label{fig:ecc-cmp}
\end{figure}

Fig.~\ref{fig:ecc-cmp} shows both the frame-error rate (left) and the bit-error rate (right) of two different $(2048, 1365)$ polar codes. The black-solid curve is the performance of a polar code constructed using the method described in \cite{Tal2011a} for $E_b/N_0 = 4$~dB. The dashed-red curve is for the $(2048, 1365)$ constructed by concatenating (assembling) a $(1024, 512)$ polar code and a $(1024, 853)$ polar code. Both polar codes of length $1024$ were also constructed using the method of \cite{Tal2011a} for $E_b/N_0$ values of 2.5 and 5 dB, respectively.

From the figure, it can be seen that constructing an optimized polar code of length 2048 with rate $\nicefrac{2}{3}$ results in a coding gain of approximately 0.17 dB at a FER of $10^{-3}$---an FER appropriate for certain applications---over one assembled from two shorter polar codes of length 1024. The gap is increasing with the signal-to-noise ratio, reaching 0.24 dB at a FER of $10^{-4}$. Looking at the BER curves, it can be observed that the gap is much narrower. Compared to that of the assembled master code, the optimized polar code shows a coding gain of 0.07 dB at a BER of $10^{-5}$.

\subsection{About Constituent Codes: frozen bit locations, rate and practicality}

The location of the frozen bits in non-optimized constituent codes is dictated by their parent code. In other words, if the master code of length $N$ has been assembled from two optimized (constituent) polar codes of length $\nicefrac{N}{2}$ as suggested in the previous section, the shorter optimized codes of length $\nicefrac{N}{2}$ determine the location of the frozen bits in their respective constituent codes of length $< \nicefrac{N}{2}$. Otherwise, the master code dictates the frozen bit locations for all constituent codes.

Assuming that the decoding algorithm takes advantage of the a priori knowledge of these locations, the code rate and frozen bit locations of constituent codes cannot be changed at execution time. However, there are many constituent codes to choose from and code shortening can be used \cite{Li2015} to create more, e.g. in order to obtain a specific number of information bits or code rate.

Because of the polarization phenomenon, given any two sibling constituent codes, the code rate of the LHS one is always lower than that of the RHS one for a properly constructed polar code \cite{Sarkis_TCOMM_2015}. That property plays to our advantage as, in many wireless applications, it is desirable to offer a variety of codes of both high and low rates.

It should be noted that not all constituent codes within a master code are of practical use e.g. codes of very high rate offer negligible coding gain over an uncoded communication. For example, among the four constituent codes of length 4 included in the $(16, 12)$ polar code illustrated in Fig.~\ref{fig:tree-sc-16-12}, two of them are rate-1 constituent codes. Using them would be equivalent to uncoded communication. Moreover, among constituent codes of the same length, many codes may have a similar number of information bits with little to no error-correction performance difference in the region of interest.

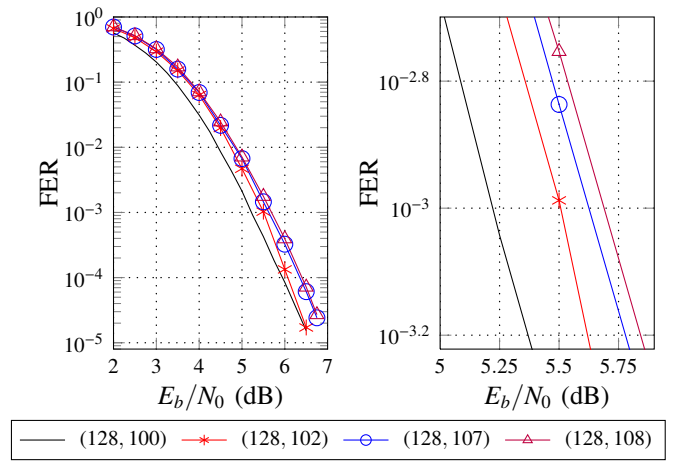
\begin{figure}
  \centering
  \definecolor{deepgreen}{RGB}{8, 130, 25}
\definecolor{mauve}{RGB}{152, 57, 153}
\definecolor{hotpink}{RGB}{255, 0, 255}

\begin{tikzpicture}

  \pgfplotsset{
    grid style = {
      dash pattern = on 0.05mm off 1mm,
      line cap = round,
      black,
      line width = 0.5pt
    },
    label style = {font=\fontsize{10pt}{7.2}\selectfont},
    tick label style = {font=\fontsize{8pt}{7.2}\selectfont}
  }

  \begin{semilogyaxis}[%
    xlabel=$E_b/N_0$ (dB),%
    xlabel style={yshift=0.5em},%
    ylabel=FER, ylabel style={yshift=-1.05em},%
    xtick={2,3,4,5,6,7},%
    xmin=2,xmax=7,ymin=8e-6,ymax=1,%
    width=0.5\columnwidth, height=6.0cm, grid=major,%
    legend style={
      anchor={center},
      cells={anchor=west},
      column sep= 1.2mm,
      font=\fontsize{8pt}{7.2}\selectfont,
    },
    legend to name=perf-2-legend,
    legend columns=4,
    mark size=3.0pt,
    mark options=solid]

    \addplot[color=black] table[x=ebn0_db,y=FER] {data/merged_n1k_awgn_s0.75+n1k_awgn_s0.436.subcode.128.100.csv};
    \addlegendentry{$(128, 100)$}

    \addplot[color=red,mark=asterisk] table[x=ebn0_db,y=FER] {data/merged_n1k_awgn_s0.75+n1k_awgn_s0.436.subcode.128.102.csv};
    \addlegendentry{$(128, 102)$}

    \addplot[color=blue,mark=o] table[x=ebn0_db,y=FER] {data/merged_n1k_awgn_s0.75+n1k_awgn_s0.436.subcode.128.107.csv};
    \addlegendentry{$(128, 107)$}

    \addplot[color=purple,mark=triangle] table[x=ebn0_db,y=FER] {data/merged_n1k_awgn_s0.75+n1k_awgn_s0.436.subcode.128.108.csv};
    \addlegendentry{$(128, 108)$}

  \end{semilogyaxis}
\end{tikzpicture}
\begin{tikzpicture}

  \pgfplotsset{
    grid style = {
      dash pattern = on 0.05mm off 1mm,
      line cap = round,
      black,
      line width = 0.5pt
    },
    label style = {font=\fontsize{10pt}{7.2}\selectfont},
    tick label style = {font=\fontsize{8pt}{7.2}\selectfont}
  }

  \begin{semilogyaxis}[%
    xlabel=$E_b/N_0$ (dB),%
    xlabel style={yshift=0.5em},%
    ylabel=FER, ylabel style={yshift=-0.8em},%
    xtick={5,5.25,5.5,5.75,6},%
    xmin=5,xmax=5.9,ymin=6e-4,ymax=2e-3,%
    width=0.5\columnwidth, height=6.0cm, grid=major,%
    mark size=3.0pt,
    mark options=solid]

    \addplot[color=black] table[x=ebn0_db,y=FER] {data/merged_n1k_awgn_s0.75+n1k_awgn_s0.436.subcode.128.100.csv};

    \addplot[color=red,mark=asterisk] table[x=ebn0_db,y=FER] {data/merged_n1k_awgn_s0.75+n1k_awgn_s0.436.subcode.128.102.csv};

    \addplot[color=blue,mark=o] table[x=ebn0_db,y=FER] {data/merged_n1k_awgn_s0.75+n1k_awgn_s0.436.subcode.128.107.csv};

    \addplot[color=purple,mark=triangle] table[x=ebn0_db,y=FER] {data/merged_n1k_awgn_s0.75+n1k_awgn_s0.436.subcode.128.108.csv};

  \end{semilogyaxis}
\end{tikzpicture}
\\
\ref{perf-2-legend}
  \caption{Error-correction performance of the four constituent codes of length 128 with a rate of approximately $\nicefrac{5}{6}$ contained in the proposed $(2048, 1365)$ master code.}
  \label{fig:ecc-cmp-128}
\end{figure}

Fig.~\ref{fig:ecc-cmp-128} shows the frame-error rate of all four constituent codes of length $128$ with a rate of approximately $\nicefrac{5}{6}$ that are contained within the proposed $(2048, 1365)$ master code. It can be seen that, even at such a short length, at a FER of $10^{-3}$ the gap between both extremes is under 0.5 dB. Among those constituent codes, only the $(128, 108)$ was selected for the implementation presented in Section~\ref{sec:results}. It is beneficial to limit the number of codes supported in a practical implementation of a multi-mode decoder in order to minimize routing circuitry.

\subsection{Latency and Throughput Considerations}

If a decoding algorithm taking advantage of the a priori knowledge of the frozen bit locations is used in the unrolled decoder, such as Fast-SSC~\cite{Sarkis_JSAC_2014}, the latency will vary even among constituent codes of the same length. However, the coded throughput will not. The coded throughput of an unrolled decoder for a polar code of length $N$ will be twice that of a constituent code of $\nicefrac{N}{2}$, which in turn, is double that of a constituent code of length $\nicefrac{N}{4}$, and so on. The coded and information throughput are defined by \eqref{eqn:tp}.

In wireless communication standards where multiple code lengths and rates are supported, the peak information throughput is typically achieved with the longest code that has both the greatest latency and highest code rate. It is not mandatory to reproduce this with our proposed method,  but it can be done if considered desirable. It is the example that we provide in the implementation section of this paper.

Another possible scenario would be to use a low-rate master code, e.g. $R=\nicefrac{1}{3}$, that is more powerful in terms of error-correction performance. The resulting multi-mode decoder would reach its peak information throughput with the longest constituent code of length $\nicefrac{N}{2}$ that has the highest code rate, a code with a significantly lower decoding latency than that of the master code.

\section{Implementation and Results}\label{sec:results}

In this section, we start by presenting results for dedicated unrolled decoders: showing the effect of the initiation interval, the code length and the code rate on unrolled decoders. Then, we present results for two implementations of our proposed multi-mode unrolled decoders. For the latter, we had the objective of building decoders with a throughput in the vicinity of 20~Gbps.

The multi-mode decoder examples are built around $(1024, 853)$ and $(2048, 1365)$ master codes. In the following, the former is referred to as the decoder supporting a maximum code length $N_{\max}$ of $1024$ and the latter as the decoder with $N_{\max}=2048$. A total of ten polar codes were selected for the decoder supporting codes of lengths up to 2048. The other decoder with $N_{\max}=1024$ has eight modes corresponding to a subset of the ten polar codes supported by the bigger decoder.
The master codes used in this section are the same as those used in Section~\ref{sec:assembled}.

For the decoder with $N_{\max}=1024$, the Repetition and SPC nodes were constrained to a maximum size $N_v$ of 8 and 4, respectively. For the decoder with $N_{\max}=2048$, we found it more beneficial to lower the execution frequency and increase the maximum sizes of the Repetition and SPC nodes to 16 and 8, respectively. Additionally, the decoder with $N_{\max}=2048$ also uses RepSPC~\cite{Sarkis_JSAC_2014} nodes to reduce latency.

\subsection{Methodology}
In our experiments, decoders are built with sufficient memory to accommodate storing an extra frame at the input, and to preserve an estimated codeword at the output. As a result, the next frame can be loaded while a frame is being decoded. Similarly, an estimated codeword can be read while the next frame is being decoded. We define decoding latency to include the time required to load channel LLRs, decode a frame and offload the estimated codeword.

\begin{figure}
  \centering
  \begin{tikzpicture}

  \pgfplotsset{
    grid style = {
      dash pattern = on 0.05mm off 1mm,
      line cap = round,
      black,
      line width = 0.5pt
    },
    label style = {font=\fontsize{9pt}{7.2}\selectfont},
    tick label style = {font=\fontsize{7pt}{7.2}\selectfont}
  }

  \begin{semilogyaxis}[%
    xlabel=$E_b/N_0$ (dB),%
    xlabel style={yshift=0.8em},%
    ylabel=FER, ylabel style={yshift=-1.05em},%
    width=0.5\columnwidth, height=6.0cm, grid=major,%
    legend style={
      anchor={center},
      cells={anchor=west},
      column sep= 2mm,
      font=\fontsize{7pt}{7.2}\selectfont,
    },
    legend to name=perf-3-legend,
    legend columns=4,
    mark size=3.0pt]

    \addplot[color=blue,mark=pentagon] table[x=ebn0_db,y=FER] {data/1024.512.float.csv};
    \addlegendentry{Float}

    \addplot[color=red,mark=triangle] table[x=ebn0_db,y=FER] {data/1024.512.fixed.6.5.1.csv};
    \addlegendentry{6.5.1}

    \addplot[color=black,mark=+] table[x=ebn0_db,y=FER] {data/1024.512.fixed.5.4.0.csv};
    \addlegendentry{5.4.0}

    \addplot[color=purple,mark=x] table[x=ebn0_db,y=FER] {data/1024.512.fixed.5.4.1.csv};
    \addlegendentry{5.4.1}

  \end{semilogyaxis}
\end{tikzpicture}
\begin{tikzpicture}

  \pgfplotsset{
    grid style = {
      dash pattern = on 0.05mm off 1mm,
      line cap = round,
      black,
      line width = 0.5pt
    },
    label style = {font=\fontsize{9pt}{7.2}\selectfont},
    tick label style = {font=\fontsize{7pt}{7.2}\selectfont}
  }

  \begin{semilogyaxis}[%
    xlabel=$E_b/N_0$ (dB),%
    xlabel style={yshift=0.8em},%
    ylabel=BER, ylabel style={yshift=-1.05em},%
    ymax=7e-2, ymin=9e-8,
    width=0.5\columnwidth, height=6.0cm, grid=major,%
    mark size=3.0pt]

    \addplot[color=blue,mark=pentagon] table[x=ebn0_db,y=BER] {data/1024.512.float.csv};

    \addplot[color=red,mark=triangle] table[x=ebn0_db,y=BER] {data/1024.512.fixed.6.5.1.csv};

    \addplot[color=black,mark=+] table[x=ebn0_db,y=BER] {data/1024.512.fixed.5.4.0.csv};

    \addplot[color=purple,mark=x] table[x=ebn0_db,y=BER] {data/1024.512.fixed.5.4.1.csv};

  \end{semilogyaxis}
\end{tikzpicture}
\\
\ref{perf-3-legend}
  \caption{Effect of quantization on the error-correction performance of a (1024, 512) polar code.}
  \label{fig:perf_quant}
\end{figure}
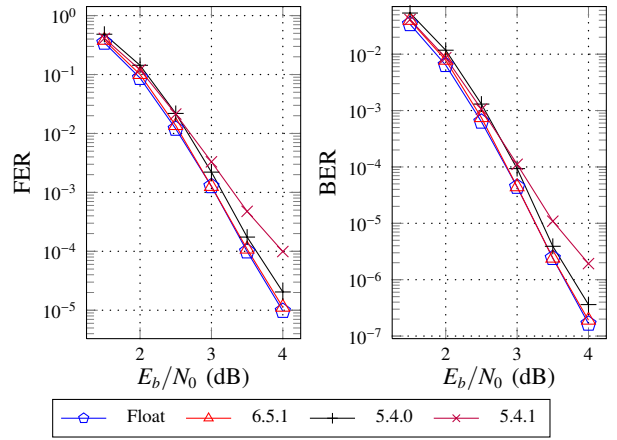

The quantization used was determined by running fixed-point simulations with bit-true models of the decoders. A smaller number of bits is used to store the channel LLRs compared to that of the other LLRs used in the decoder. All LLRs use 2's complement representation and share the same number of fractional bits. We denote quantization as $Q_i$.$Q_c$.$Q_f$, where $Q_c$ is the total number of bits to store a channel LLR, $Q_i$ is the total the number of bits used to store internal LLRs and $Q_f$ is the number of fractional bits in both. $Q_i$ and $Q_c$ both include the sign bit. Fig.~\ref{fig:perf_quant} shows that, for a (1024, 512) polar code modulated with BPSK and transmitted over an AWGN channel, using $Q_i$.$Q_c$.$Q_f$ equal to $5.4.0$ results in a 0.1 dB performance degradation at a bit-error rate of $10^{-6}$. Thus we used that quantization for the hardware results. 

ASIC synthesis results are for the 65~nm CMOS GP technology from TSMC and are obtained with Cadence RTL Compiler. Unless indicated otherwise, all results are for the worst-case library at a supply voltage of 0.72 V with an operating temperature of $125^\circ$C. Power consumption estimations are also obtained from Cadence RTL Compiler, switching activity is derived from simulation vectors.
Only registers were used for memory due to the lack of access to an SRAM compiler.

\subsection{Dedicated Decoders: Effect of the Initiation Interval}
In this section, we explore the effect of the initiation interval on the implementation of the fully-unrolled architecture. The decoders are built for the same (1024, 512) polar code used in \cite{Giard_IET_2015}, although many improvements were made since the publication of that work. Regardless of the initiation interval, all decoders use 5.4.0 quantization and have a decoding latency of 364 clock cycles.

\begin{table}[t]
  \centering
  \setlength{\tabcolsep}{4pt}
  \caption{Decoders for a (1024, 512) polar code with various initiation intervals $\mathcal{I}$. The clock is set to 500~MHz and the latency is of 728 ns.}
  \begin{tabular}{ccccrrc}
    \toprule
    \multirow{2}{*}{$\mathcal{I}$} & \textbf{Tot. Area} & \textbf{Log. Area} & \textbf{Mem. Area} & \multicolumn{1}{c}{\textbf{T/P}} & \multicolumn{1}{c}{\textbf{Power}} & \textbf{Energy} \\
    & {\footnotesize(mm$^2$)} & {\footnotesize(mm$^2$)} & {\footnotesize(mm$^2$)} & {\footnotesize(Gbps)}& {\footnotesize(mW)} & {\footnotesize(pJ/bit)}\\
    \midrule
    1 &           12.369 & 0.60 &           11.75 & 512.0           & 3,830 & \phantom{0}7.5\\ 
    4 & \phantom{0}4.921 & 0.64 & \phantom{0}4.24 & 128.0           & 1,060 & \phantom{0}8.3 \\ 
   50 & \phantom{0}1.232 & 0.65 & \phantom{0}0.56 & \phantom{0}10.2 &   107 & 10.5\\ 
  167 & \phantom{0}0.998 & 0.63 & \phantom{0}0.34 & \phantom{00}3.1 &    62 & 20.0\\ 
    \bottomrule
  \end{tabular}
  \label{tab:impl:asic:initiation_interval}
\end{table}

Table~\ref{tab:impl:asic:initiation_interval} shows the results for various initiation intervals. Besides the effect on throughput, increasing the initiation interval causes a significant reduction in memory requirements without significantly affecting combinational logic.
Since area is largely dominated by registers, increasing the initiation interval has great effect on the total area. For example, using $\mathcal{I}=50$ results in an area that is more than 10 times smaller, at the cost of a throughput that is 50 times lower. That table also shows that reducing the area has a direct effect on the estimated power consumption, which significantly drops as $\mathcal{I}$. 

As expected, increasing the initiation interval $\mathcal{I}$ offers a diminishing return as it gets closer to the maximum, 167 for the example (1024, 512) code. Also, as $\mathcal{I}$ is increased, the energy efficiency is reduced.

\subsection{Dedicated Decoders: Effect of the Code Length and Rate}\label{subsec:codelength}
Results for other polar codes are presented in this section where we show the effect of the code length and rate on performance and resource usage.

\begin{table}[h]
  \centering
  \setlength{\tabcolsep}{2pt}
  \caption{Deeply-pipelined decoders for polar codes of various lengths with rate $R=\nicefrac{1}{2}$. The clock is set to 500~MHz.}
  \begin{tabular}{cccccrrc}
    \toprule
    \multirow{2}{*}{$N$} & \textbf{Tot. Area} & \textbf{Log. Area} & \textbf{Mem. Area} & \textbf{Latency} & \multicolumn{1}{c}{\textbf{T/P}} & \multicolumn{1}{c}{\textbf{Power}} & \multicolumn{1}{c}{\textbf{Energy}}\\
    & {\footnotesize(mm$^2$)} & {\footnotesize(mm$^2$)} & {\footnotesize(mm$^2$)} & {\footnotesize(ns)} & {\footnotesize(Gbps)} & {\footnotesize(mW)} & {\footnotesize(pJ/bit)}\\
    \midrule
   128 & \phantom{0}0.349 & 0.05 & \phantom{0}0.29 & 152 &  64 &   105 & \phantom{0}1.6\\
   256 & \phantom{0}1.121 & 0.12 & \phantom{0}0.99 & 268 & 128 &   342 & \phantom{0}2.7\\
   512 & \phantom{0}3.413 & 0.27 & \phantom{0}3.14 & 408 & 256 & 1,050 & \phantom{0}4.0\\
  1024 &           12.369 & 0.60 &           11.75 & 728 & 512 & 3,830 & \phantom{0}7.5\\
  2048 &           43.541 & 1.32 &           42.16 &1,304&1,024&13,526 &13.2\\
    \bottomrule
  \end{tabular}
  \label{tab:impl:asic:code_length}
\end{table}

Tables~\ref{tab:impl:asic:code_length} and \ref{tab:impl:asic:code_length_partial} show the effect of the code length on area, decoding latency, coded throughput, power consumption, and on energy efficiency for polar codes of short to moderate lengths. Table~\ref{tab:impl:asic:code_length} contains results for the fully-unrolled deeply-pipelined architecture ($\mathcal{I}=1$) and the code rate $R$ is fixed to \nicefrac{1}{2} for all polar codes. Table~\ref{tab:impl:asic:code_length_partial} contains results for the fully-unrolled partially-pipelined architecture where the maximum initiation interval ($\mathcal{I}_{\max}$) is used and the code rate $R$ is \nicefrac{5}{6}.

As shown in Table~\ref{tab:impl:asic:code_length}, with a deeply-pipelined architecture, logic area usage almost grows as $N\log_2N$, whereas memory area is closer to being quadratic in code length $N$. The logic area required for a deeply-pipelined unrolled decoder implemented in 65~nm ASIC technology can be approximated with an accuracy greater than 98\%  using $C \cdot N\log_{2}{}N$, where the constant $C$ is set to $\nicefrac{1}{17,000}$. For comparison, the logic area of tree-based SC decoders is $\mathcal{O}(N)$ while the other state-of-the-art partially-parallel architectures have fixed logic area that do not depend on the code length.

Curve fitting shows that the memory area is quadratic with code length $N$. Let the memory area be defined by $a+bN+cN^2$, setting $a=0.249$, $b=2.466\times10^{-3}$ and $c=8.912\times10^{-6}$ results in a standard error of 0.1839.

As shown in Table~\ref{tab:impl:asic:code_length}, throughput exceeding 1 Tbps and 500 Gbps can be achieved with a deeply-pipelined decoder for polar codes of length $2048$ and $1024$, respectively. As the memory area grows quadratically with the code length the amount of energy required to decode a bit increases with the code length.
The decoder for the $(4096, 2048)$ polar code could not be synthesized on our server due to insufficient memory.

\begin{table}[h]
  \centering
  \setlength{\tabcolsep}{3pt}
  \caption{Partially-pipelined decoders with initiation interval set to $\mathcal{I}_{\max}$ for polar codes of various lengths with rate $R=\nicefrac{5}{6}$. The clock is set to 500~MHz.}
  \begin{tabular}{cccccccc}
    \toprule
    \multirow{2}{*}{$N$} & \multirow{2}{*}{$\mathcal{I}$} & \textbf{Tot. Area} & \textbf{Mem. Area} & \textbf{Latency} & \textbf{T/P} & \textbf{Power} & \textbf{Energy}\\
    & & {\footnotesize(mm$^2$)} & {\footnotesize(mm$^2$)} & \footnotesize($\mu$s) & \footnotesize(Gbps) & \footnotesize(mW) & \footnotesize(pJ/bit)\\
    \midrule
    1024 & 206 & 0.793 & 0.28 & 0.646 & 2.5 & \phantom{0}51 & 20.5\\
    2048 & 338 & 1.763 & 0.61 & 0.888 & 3.0 &           108 & 35.6\\
    4096 & 665 & 4.248 & 1.44 & 1.732 & 3.1 &           251 & 81.5\\
    \bottomrule
  \end{tabular}
  \label{tab:impl:asic:code_length_partial}
\end{table}

For a partially-pipelined architecture with $\mathcal{I}_{\max}$, both the memory and total area scale linearly with $N$. The power consumption is shown to almost scale linearly as well. The results of  Table~\ref{tab:impl:asic:code_length_partial} also show that it was possible to synthesize ASIC decoders for larger code lengths than what was possible with a deeply-pipelined architecture.

\begin{table}[h]
  \centering
  \caption{Deeply-pipelined decoders for polar codes of length $N=1024$ with common rates. The clock is set to 500~MHz and the throughput is of 512 Gbps.}
  \begin{tabular}{ccccccc}
    \toprule
    \multirow{3}{*}{$R$} & \multirow{3}{*}{\shortstack{\textbf{Tot. Area}\\\footnotesize(mm$^2$)}} & \multirow{3}{*}{\shortstack{\textbf{Mem. Area}\\\footnotesize(mm$^2$)}} & \multicolumn{2}{c}{\textbf{Latency}} & \multirow{3}{*}{\shortstack{\textbf{Power}\\\footnotesize(mW)}} & \multirow{3}{*}{\shortstack{\textbf{Energy}\\\footnotesize(pJ/bit)}}\\
    \cmidrule(lr){4-5}
    & & & (CCs) &\footnotesize(ns) & & \\
    \midrule
    \nicefrac{1}{2} & 12.369 & 11.75 & 364 & 727 & 3,830 & 7.5\\
    \nicefrac{2}{3} & 13.049 & 12.45 & 326 & 651 & 4,041 & 6.2\\
    \nicefrac{3}{4} & 15.676 & 15.05 & 373 & 745 & 4,865 & 6.5\\
    \nicefrac{5}{6} & 14.657 & 14.05 & 323 & 645 & 4,549 & 7.1\\
    \bottomrule
  \end{tabular}
  \label{tab:impl:asic:code_rate}
\end{table}

The effect of using different code rates for a polar code of length $N=1024$ is shown in Table~\ref{tab:impl:asic:code_rate}. 
We note that the higher rate codes do not have noticeably lower latency compared to the rate-$\nicefrac{1}{2}$ code, contrary to what was observed in \cite{Sarkis_JSAC_2014}. This is due to limiting the width of SPC nodes to $N_{\text{SPC}}=4$ in this work, whereas it was left unbounded in the others. The result is that long SPC codes are implemented as trees whose left-most child is a width-4 SPC node and the others are all rate-1 nodes.
Thus, for each additional stage $(\log_2N_v-\log_2N_{\text{SPC}})$ of an SPC code of length $N_v > N_{\text{SPC}}$, four nodes with a total latency of 3 clock cycles are required: $F$, $G$ followed by $I$, and $Combine$. This brings the total latency of decoding a long SPC code to $3(\log_2N_v-\log_2N_{\text{SPC}})+1$ clock cycles compared to $\left \lceil \nicefrac{N_v}{\mathcal{P}}\right \rceil + 4$ in \cite{Sarkis_JSAC_2014}, where $\mathcal{P}$ is the number of LLRs that can be read simultaneously (256 was a typical value for $\mathcal{P}$ in \cite{Sarkis_JSAC_2014}).

From Table~\ref{tab:impl:asic:code_rate}, it can be seen that varying the rate does not affect the logic area that remains almost constant at approximately 0.61 mm$^2$. Memory, in the form of registers, dominates the decoder area. Therefore, the estimated power consumption scales according to the memory area.

\subsection{Deeply-pipelined SC Decoders}
To decode a frame, an SC decoder needs to load a frame, visit all $\sum_{i = 1}^{\log_2N} 2^i$ edges of the decoder tree twice and store the estimated codeword. A deeply-pipelined SC decoder for a (128, 64) polar code has an area of 2.17 mm$^2$, a latency of 510 clock cycles, and a power consumption of 677 mW. These values are 6.2, 6.7, and 6.4 times as much as their counterparts of the deeply-pipelined Fast-SSC decoder reported in Table~\ref{tab:impl:asic:code_length}. These results indicate that deeply-pipelined SC decoders will be limited to very short polar codes, and that alternative algorithms and architectures will yield more practical implementations.

\subsection{Multi-mode Decoders: Error-correction Performance}
\begin{figure}
  \centering
  \definecolor{deepgreen}{RGB}{8, 130, 25}
\definecolor{mauve}{RGB}{152, 57, 153}
\definecolor{hotpink}{RGB}{255, 0, 255}

\begin{tikzpicture}

  \pgfplotsset{
    grid style = {
      dash pattern = on 0.05mm off 1mm,
      line cap = round,
      black,
      line width = 0.5pt
    },
    label style = {font=\fontsize{10pt}{7.2}\selectfont},
    tick label style = {font=\fontsize{8pt}{7.2}\selectfont}
  }

  \begin{semilogyaxis}[%
    xlabel=$E_b/N_0$ (dB),%
    xlabel style={yshift=0.5em},%
    ylabel=FER, ylabel style={yshift=-1.05em},%
    xtick={2,3,4,5,6,7,8},%
    xmin=2,xmax=8,ymin=8e-6,ymax=1,%
    width=0.75\columnwidth, height=6.5cm, grid=major,%
    legend style={
      at={(1.475,0.9)},
      column sep= 1mm,
      font=\fontsize{8pt}{7.2}\selectfont,
    },
    legend columns=1,
    mark size=3.0pt,
    mark options=solid]

    \addplot[color=black,dashed] table[x=ebn0_db,y=FER] {data/merged_n1k_awgn_s0.75+n1k_awgn_s0.436.float.csv};
    \addlegendentry{$(2048, 1365)$}

    \addplot[color=red,dashed,mark=asterisk] table[x=ebn0_db,y=FER] {data/1k_s0.75_syst.csv};
    \addlegendentry{$(1024, 512)$}

    \addplot[color=black] table[x=ebn0_db,y=FER] {data/1024.853.s0.436.int8.csv};
    \addlegendentry{$(1024, 853)$}

    \addplot[color=red,mark=triangle] table[x=ebn0_db,y=FER] {data/n1k_awgn_s0.436.subcode.512.490.csv};
    \addlegendentry{$(512, 490)$}

    \addplot[color=blue,mark=+] table[x=ebn0_db,y=FER] {data/n1k_awgn_s0.436.subcode.512.363.csv};
    \addlegendentry{$(512, 363)$}

    \addplot[color=purple,mark=x] table[x=ebn0_db,y=FER] {data/n1k_awgn_s0.436.subcode.256.228.csv};
    \addlegendentry{$(256, 228)$}

    \addplot[color=deepgreen,mark=o] table[x=ebn0_db,y=FER] {data/n1k_awgn_s0.436.subcode.256.135.csv};
    \addlegendentry{$(256, 135)$}

    \addplot[color=orange,mark=diamond] table[x=ebn0_db,y=FER] {data/n1k_awgn_s0.436.subcode.128.108.csv};
    \addlegendentry{$(128, 108)$}

    \addplot[color=hotpink,mark=pentagon] table[x=ebn0_db,y=FER] {data/n1k_awgn_s0.436.subcode.128.96.csv};
    \addlegendentry{$(128, 96)$}

    \addplot[color=cyan,mark=star] table[x=ebn0_db,y=FER] {data/n1k_awgn_s0.436.subcode.128.39.csv};
    \addlegendentry{$(128, 39)$}

  \end{semilogyaxis}
\end{tikzpicture}
  \vspace{-5pt}
  \caption{Error-correction performance of the polar codes.}
  \label{fig:ecc-perf}
\end{figure}
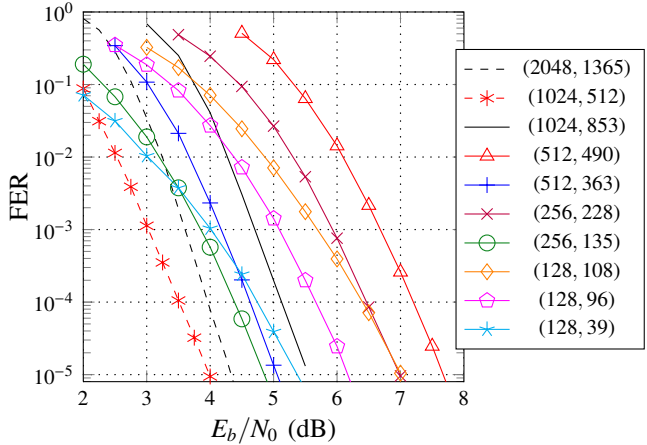

Fig.~\ref{fig:ecc-perf} shows the frame-error rate performance of ten different polar codes. The decoder with $N_{\max}=2048$ supports all ten illustrated polar codes whereas the decoder with $N_{\max}=1024$ supports all polar codes but the two shown as dotted curves. All simulations are generated using random codewords modulated with binary phase-shift keying and transmitted over an additive white Gaussian channel.

It can be seen from the figure that the error-correction performance of the supported polar codes varies greatly. As expected, for codes of the same lengths, the codes with the lowest code rates performs significantly better than their higher rate counterpart. For example, at a FER of $10^{-4}$, the performance of the $(512, 363)$ polar code is almost 3 dB better than that of the $(512, 490)$ code.

While the error-correction performance plays a role in the selection of a code, the latency and throughput are also important considerations. As it will be shown in the following section, the ten selected polar codes perform much differently in that regard as well.

\subsection{Multi-mode Decoders: Latency and Throughput}

Table~\ref{tab:tp_and_latency_results} shows the latency and information throughput for both decoders with $N_{\max}\in\{1024,2048\}$. To reduce the area and latency while retaining the same throughput, the initiation interval $\mathcal{I}$ can be increased along with the clock frequency~\eqref{eqn:tp}.

If both decoders have initiation intervals of $20$---as used in the section below---Table~\ref{tab:tp_and_latency_results} assumes clock frequencies of 500~MHz and 250~MHz for the decoders with $N_{\max}=1024$ and $N_{\max}=2048$, respectively. While their master codes differ, both decoders feature a peak information throughput in the vicinity of 20 Gbps. For the decoder with the smallest $N_{\max}$, the seven other polar codes have an information throughput in the multi-gigabit per second range with the exception of the shortest and lowest-rate constituent code. That $(128, 39)$ constituent code still has an information throughput close to 1 Gbps. The decoder with $N_{\max}=2048$ offers multi-gigabit throughput for most of the supported polar codes. The minimum information throughput is also with the $(128, 39)$ polar code at approximately 500~Mbps.

\begin{table}
  \centering
  \setlength{\tabcolsep}{0.65pt}
  \caption{Information throughput and latency for the multi-mode unrolled polar decoders based on the $(2048, 1365)$ and $(1024, 853)$ master codes, respectively with a $N_{\max}$ of 1024 and 2048.}
  \begin{tabular}{c c C{0.7cm} *{6}{C{0.9cm}}}
    \toprule
    \multirow{2}{*}{\Shortunderstack{\textbf{Code}\\$(N, k)$}} & \multirow{2}{*}{\Shortunderstack{\textbf{Rate}\\$(\nicefrac{k}{N})$}} && \multicolumn{2}{c}{\textbf{Info. T/P} (Gbps)} & \multicolumn{2}{c}{\textbf{Latency} (CCs)} & \multicolumn{2}{c}{\textbf{Latency} (ns)}\\
    \cmidrule(lr){4-5}\cmidrule(lr){6-7}\cmidrule(lr){8-9}
    &&$N_{\max}=$& 1024 & 2048 & 1024 & 2048 & 1024 & 2048\\
    \midrule
    (2048, 1365)& 2/3  &&    -  & 17.1 &   - & 503 & -   & 2,012\\
    (1024, 853) & 5/6  && 21.3  & 10.7 & 323 & 236 & 646 &   944\\
    (1024, 512) & 1/2  &&    -  &  6.4 &   - & 265 & -   & 1,060\\
    (512, 490)  &19/20 && 12.3  &  6.2 &  95 & 75  & 190 &   300\\
    (512, 363)  & 7/10 &&  9.1  &  4.5 & 226 & 159 & 452 &   636\\
    (256, 228)  & 9/10 &&  5.7  &  2.6 &  86 & 61  & 172 &   244\\
    (256, 135)  & 1/2  &&  3.4  &  1.7 & 138 & 96  & 276 &   384\\
    (128, 108)  & 5/6  &&  2.7  &  1.4 &  54 & 40  & 108 &   160\\
    (128, 96)   & 3/4  &&  2.4  &  1.2 &  82 & 52  & 164 &   208\\
    (128, 39)   & 1/3  &&  0.98 &  0.49&  54 & 42  & 108 &   168\\
    \bottomrule
  \end{tabular}
  \label{tab:tp_and_latency_results}
\end{table}

In terms of latency, the decoder with $N_{\max}=1024$ requires $646$~ns to decode its longest supported code. The latency for all the other codes supported by that decoder is under $500$~ns. Even with its additional dedicated node and relaxed maximum size constraint on the Repetition and SPC nodes, the decoder with $N_{\max}=2048$ has greater latency overall because of its lower clock frequency. For example, its latency is of 2.01~$\mu$s, 944~ns and 1.06~$\mu$s for the $(2048, 1365)$, $(1024,853)$ and $(1024,512)$ polar codes, respectively.

Using the same nodes and constraints as for $N_{\max}=1024$, the $N_{\max}=2048$ decoder would allow for greater clock frequencies. While 689 clocks cycles would be required to decode the longest polar code instead of 503, a clock of 500~MHz would be achievable, effectively reducing the latency from 2.01~$\mu$s to 1.38~$\mu$s and doubling the throughput. However, this reduction comes at the cost of much greater area and an estimated power consumption close to 1~W.

\subsection{Comparing with the State of the Art}
\begin{table*}
  \centering
  \caption{Comparison with state-of-the-art polar decoders.}
  \begin{tabular}{l cccc c c c c c}
    \toprule
                 & \multicolumn{4}{c}{\bf Multi-mode}& {\bf Dedicated} & \cite{Giard_JSPS_2016} & \phantom{$^\diamond$}\cite{Park2014}$^\diamond$& \cite{Dizdar2015} & \cite{Yuan2014} \\
    \midrule
    \textbf{Algorithm}    & \multicolumn{4}{c}{Fast-SSC} & Fast-SSC             & Fast-SSC & BP     & SC        & 2-bit SC \\
    \textbf{Technology}   &  \multicolumn{4}{c}{65 nm}   & 65 nm                & 65 nm    & 65 nm  & 90 nm     & 45 nm \\
    $\bm{N_{\max}}$         & \multicolumn{2}{c}{1024} &\multicolumn{2}{c}{2048}&1024&1024    & 1024   & 1024      & 1024 \\
    \textbf{Code}         & \multicolumn{2}{c}{$(1024,853)$} &\multicolumn{2}{c}{$(2048,1365)$}&$(1024,512)$&$(1024,512)$&$(1024,512)$& $(1024,k)$&$(1024,512)$ \\
    \textbf{Init. Interval} ($\mathcal{I}$)&20&26&    20 &   28 &  20  &    - & -            & - & - \\
    \textbf{Supply} (V)   &          0.72 & 1.0  &  0.72 &  1.0 & 1.0  &  1.0 & 1.0          & 1.3 & N/A \\   
    \textbf{Oper. temp.} ($^\circ$C)&  125 &  25  &   125 &   25 & 25   &  25  &$\approx 25$  & N/A & N/A \\ 
    \textbf{Area} (mm$^2$)&          1.71 & 1.44 &  4.29 & 3.58 & 1.68 & 0.69 & 1.48         & 3.21 & N/A \\ 
    \textbf{Area @65nm} (mm$^2$) &   1.71 & 1.44 &  4.29 & 3.58 & 1.68 & 0.69 & 1.48         & 1.68 & 0.4 \\ 
    \textbf{Frequency} (MHz)&        500  &  650 &   250 &  350 &  500 &  600 & 300          & 2.5 & 750 \\  
    \textbf{Latency} ($\mu$s)&       0.65 & 0.50 &  2.01 & 1.44 & 0.73 & 0.27 & 50           & 0.39 & 1.02 \\
    \textbf{Coded T/P} (Gbps)&       25.6 & 25.6 &  25.6 & 25.6 & 25.6 &  3.7 & 4.7 @ 4 dB   & 2.56 & 1.0 \\ 
    \textbf{Sust. Coded T/P} (Gbps)& 25.6 & 25.6 &  25.6 & 25.6 & 25.6 &  3.7 & 2.0          & 2.56 & 1.0 \\ 
    \textbf{Area Eff.} (Gbps/mm$^2$)&15.42& 17.75&  5.97 & 7.16 & 15.27& 5.40 & 3.18 @ 4 dB  & 0.80 & N/A \\
    \textbf{Power} (mW)   &          226  &  546 &   379 &  740 &  386 &  215 & 478          & 191  & N/A \\ 
    \textbf{Energy} (pJ/bit)&        8.8  & 21.3 &  14.8 & 28.9 & 15.1 & 57.7 & 102.1        & 74.5 & N/A \\ 
    \bottomrule
    &&&&&\vspace{-6pt}\\
    \multicolumn{5}{l}{\textit{$\diamond$ Measurement results.}} &\\
  \end{tabular}
  \vspace{-17pt}
  \label{tab:cmp_asic}
\end{table*}

Table~\ref{tab:cmp_asic} shows the synthesis results along with power consumption estimations for the two implementations of the proposed multi-mode unrolled decoder. The work in the first two columns is for the decoder with $N_{\max}=1024$, based on the $(1024, 853)$ master code. It was synthesized for clock frequencies of 500~MHz and 650~MHz, respectively, with initiation intervals $\mathcal{I}$ of 20 and 26. Our work shown in the third and fourth columns is for the decoders with $N_{\max}=2048$, built from the assembled $(2048, 1365)$ polar code. These decoders have an initiation interval $\mathcal{I}$ of 20 or 28, with lower clock frequencies of 250~MHz and 350~MHz, respectively. For comparison with other works, the same table also includes results for a dedicated partially-pipelined decoder for a $(1024, 512)$ polar code.

The four fastest polar decoder implementations from the literature are also included for comparison along with normalized area results. For consistency, only the largest polar code supported by each of our proposed multi-mode unrolled decoders is used and the coded throughput, as opposed to the information one, is compared to match what was done in most of the other works.

From Table~\ref{tab:cmp_asic}, it can be seen that the area for the proposed decoders with $N_{\max}=1024$ are similar to that of the BP decoder of \cite{Park2014} as well as the normalized area for the unrolled SC decoder from \cite{Dizdar2015}. However, their area is from 2.1 to 2.5 times greater than that of \cite{Giard_JSPS_2016}. Comparing the multi-mode decoders, the area for the decoder with $N_{\max}=2048$ is over twice that of the ones with $N_{\max}=1024$, however the master code for the former has twice the length of the latter and supports two more modes.

All proposed decoders have a coded throughput that is an order of magnitude greater than the other works. Latency is one to two orders of magnitude lower than that of the BP decoder. Comparing against the SC decoder of \cite{Dizdar2015}, the latency is 1.7 or 3.7 times greater for decoders with an $N_{\max}$ of 1024 and 2048, respectively. It should be noted that the decoder of \cite{Dizdar2015} support codes of any rate, where the proposed multi-mode decoders support a limited number of code rates.

The latency of the proposed decoders is higher than the programmable Fast-SSC decoder of \cite{Giard_JSPS_2016}. This is due to greater limitations on the specialized repetition and SPC decoders. The decoder in \cite{Giard_JSPS_2016} limits repetition decoders to a maximum length of 32, compared to 8 or 16 in this work, and does not place limits on the SPC decoders.

Finally, among the decoders with $N_{\max}=1024$ implemented in 65~nm with a 1~V power supply and operating at $25^\circ$C, our proposed implementation offers the greatest area and energy efficiency. The proposed multi-mode decoder exhibits 3.3 and 5.6 times better area efficiency than the decoders of \cite{Giard_JSPS_2016} and \cite{Park2014}, respectively. The energy efficiency is estimated to be 2.7 and 4.8 times higher compared to that of the same two decoders from the literature.
 
Recently, a List-based multi-mode decoder was proposed in \cite{Xiong_TVLSI_2015}, where the definition of the word ``multi-mode'' differs greatly with our work: in our work, it is used to indicate that the decoder is capable of decoding codes with varying length and rate. Whereas in \cite{Xiong_TVLSI_2015}, a ``mode'' indicates the level of parallelism in the decoder. The decoder of \cite{Xiong_TVLSI_2015} is capable of decoding 4 paths in parallel by implementing 4 processing units. It can be configured to either do SC-based decoding of 4 frames or List-based decoding. For the latter, two list sizes $L$ are supported. If $L=2$, 2 frames are decoded in parallel otherwise if $L=4$, only 1 frame is decoded at a time.

\subsection{I/O Bounded Decoding}
The family of unrolled architectures that we proposed requires tremendous throughput at the input of the decoder, especially with a deeply-pipelined architecture. For example, if a quantization of $Q_c=4$ bits is used for channel LLRs, for every estimated bit, 4 times as many bits have to be loaded into the decoder. In other words, the total data rate is 5 times that of the output. This can be a significant challenge on both FPGA and ASIC.
If only for that reason, partially-pipelined architectures are certainly more attractive.

\section{Conclusion}\label{sec:conclusion}
In this paper we presented a family of architectures for fully-unrolled polar decoders. With an initiation interval that can be adjusted, these architectures make it possible to find a trade-off between area and achievable throughput without affecting decoding latency. We showed that a fully-unrolled deeply-pipelined decoder implemented on an ASIC could achieve a throughput up to three orders of magnitude greater than the state of the art. Furthermore, we presented a new method to transform an unrolled architecture into a multi-mode decoder supporting various polar code lengths and rates. We showed that a master code can be assembled from two optimized polar codes of smaller length, with desired code rates, without sacrificing too much coding gain. We provided results for two decoders, one built for a $(1024, 853)$ master code and the other for a longer $(2048, 1365)$ polar code. Both decoders support from seven to nine other practical codes. On 65~nm ASIC, they were shown to have a peak throughput greater than 25~Gbps. One has a worst-case latency of 2~$\mu$s at 250~MHz and an energy efficiency of 14.8 pJ/bit. The other has a worst-case latency of 646~ns at 500~MHz and an energy efficiency of 8.8 pJ/bit. Both implementation examples show that, with their great throughput and support for codes of various lengths and rates, multi-mode unrolled polar decoders are promising candidates for future wireless communication standards.

\section*{ACKNOWLEDGEMENT}
Claude Thibeault is a member of ReSMiQ. Warren J. Gross is a member of ReSMiQ and SYTACom.

\bibliographystyle{IEEEtran}
\bibliography{IEEEabrv,refs}

\begin{IEEEbiography}[{\includegraphics[width=1in,height=1.25in,clip,keepaspectratio]{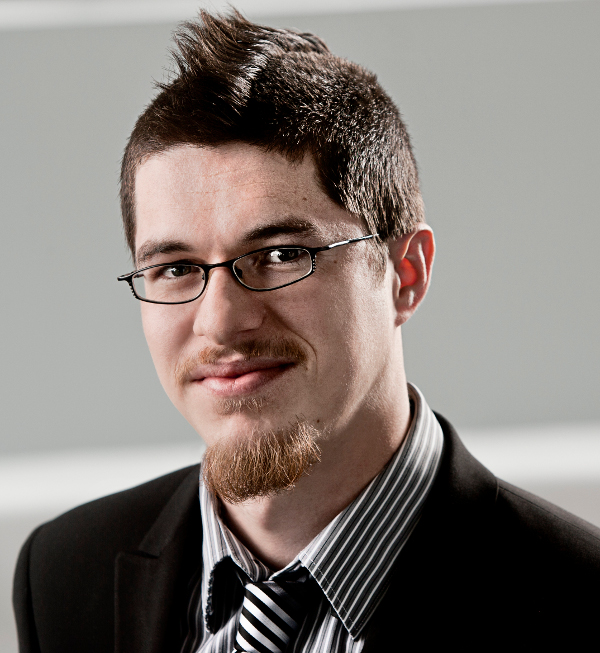}}]{Pascal Giard} received the B.Eng. and M.Eng. degree in electrical engineering from \'{E}cole de technologie sup\'erieure (\'{E}TS), Montreal, QC, Canada, in 2006 and 2009.
From 2009 to 2010, he worked as a research professional in the NSERC-Ultra Electronics Chair on 'Wireless Emergency and Tactical
Communication' at \'{E}TS.
He is currently working toward the Ph.D. degree at McGill University.
His research interests are in the design and implementation of signal processing systems with a focus on modern error-correcting codes.
\end{IEEEbiography}

\begin{IEEEbiography}[{\includegraphics[width=1in,height=1.25in,clip,keepaspectratio]{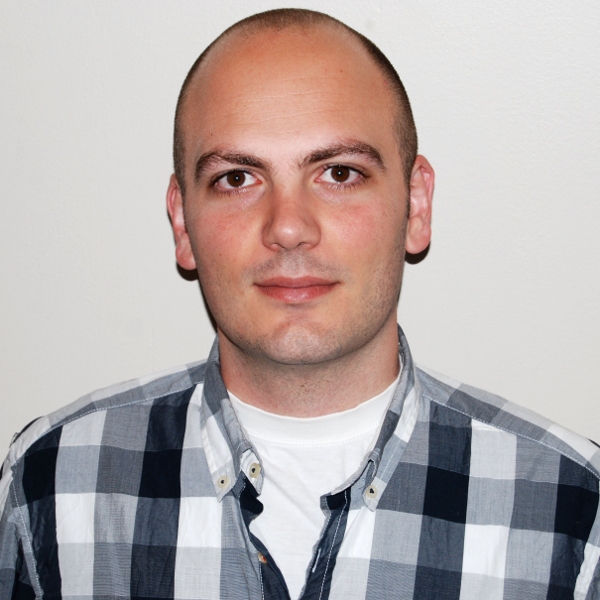}}]{Gabi Sarkis} received the B.Sc. degree in electrical engineering from Purdue University, West Lafayette, Indiana, United States, in 2006 and the M.Eng. and Ph.D. degrees from McGill University, Montreal, Quebec, Canada, in 2009 and 2016, respectively. His research interests are in the design of efficient algorithms and implementations for decoding error-correcting codes, in particular non-binary LDPC and polar codes.
\end{IEEEbiography}

\begin{IEEEbiography}[{\includegraphics[width=1in,height=1.25in,clip,keepaspectratio]{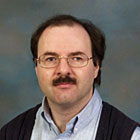}}]{Claude Thibeault}
received his Ph.D. from Ecole Polytechnique de Montreal, Canada. He is now with the Electrical Engineering department of Ecole de technologie superieure, where he serves as full professor.
His research interests include design and verification methodologies targeting ASICs and FPGAs, defect and fault tolerance, radiation effects, as well as IC and PCB test and diagnosis. He holds 13 US patents and has published more than 140 journal and conference papers, which were cited more than 850 times. He co-authored the best paper award at DVCON'05, verification category. He has been a member of different conference program committees, including the VLSI Test Symposium, for which he was program chair in 2010--2012, and general chair in 2014 and 2015.
\end{IEEEbiography}

\begin{IEEEbiography}[{\includegraphics[width=1in,height=1.25in,clip,keepaspectratio]{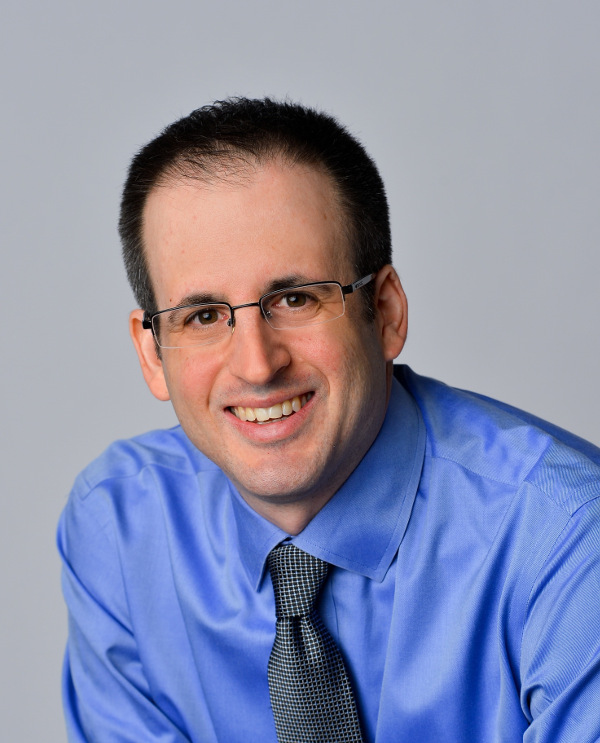}}]{Warren J. Gross}
received the B.A.Sc. degree in electrical engineering from the University of Waterloo, Waterloo, Ontario, Canada, in 1996, and the M.A.Sc. and Ph.D. degrees from the University of Toronto, Toronto, Ontario, Canada, in 1999 and 2003, respectively. Currently, he is an Associate Professor with the Department of Electrical and Computer Engineering, McGill University, Montr\'eal, Qu\'ebec, Canada. His research interests are in the design and implementation of signal processing systems and custom computer architectures.
Dr. Gross is currently Chair of the IEEE Signal Processing Society Technical Committee on Design and Implementation of Signal Processing Systems. He has served as Technical Program Co-Chair of the IEEE Workshop on Signal Processing Systems (SiPS 2012) and as Chair of the IEEE ICC 2012 Workshop on Emerging Data Storage Technologies. Dr. Gross served as Associate Editor for the IEEE Transactions on Signal Processing. He has served on the Program Committees of the IEEE Workshop on Signal Processing Systems, the IEEE Symposium on Field-Programmable Custom Computing Machines, the International Conference on Field-Programmable Logic and Applications and as the General Chair of the 6th Annual Analog Decoding Workshop. Dr. Gross is a Senior Member of the IEEE and a licensed Professional Engineer in the Province of Ontario.
\end{IEEEbiography}

\vfill 

\end{document}